\RequirePackage{ifpdf}
\documentclass[preprint, preprintnumbrs]{article}
\usepackage{jheppub}
\usepackage{xcolor}
\usepackage{amsmath}
\usepackage{epsfig}
\usepackage{caption}
\usepackage{subcaption}
\usepackage{soul}
\usepackage{cancel}
\pdfoutput=1

\newcommand{\roughly}[1]{\mathrel{\raise.3ex\hbox{$#1$\kern-0.85em
\lower1ex\hbox{$\sim$}}}}

\def\nn{\nonumber}

\newcommand{\be}{\begin{equation}}
\newcommand{\bee}{\begin{equation}}
\newcommand{\ee}{\end{equation}}
\newcommand{\beea}{\begin{eqnarray}}
\newcommand{\eea}{\end{eqnarray}}
\newcommand{\bea}{\begin{eqnarray}}

\def\nott#1{\setbox0=\hbox{$#1$}                
   \dimen0=\wd0                                 
   \setbox1=\hbox{/} \dimen1=\wd1               
   \ifdim\dimen0>\dimen1                        
      \rlap{\hbox to \dimen0{\hfil/\hfil}}      
      #1                                        
   \else                                        
      \rlap{\hbox to \dimen1{\hfil$#1$\hfil}}   
      /                                         
   \fi}                                         %

\def\uxsl{\hbox{/\kern-.4000em$u$}}
\def\uxslsm{\hbox{\smaller/\kern-.5600em$u$}}
\def\pxpsl{\hbox{/\kern-.5000em$p$}}
\def\epssl{\hbox{/\kern-.5600em$\epsilon$}}
\def\delsl{\hbox{/\kern-.7000em$\nabla$}}
\def\lxpsl{\hbox{/\kern-.5600em$l$}}
\def\kxpsl{\hbox{/\kern-.5600em$k$}}
\def\qxpsl{\hbox{/\kern-.3900em$q$}}

\def\pref#1{(\ref{#1})}
\def\exd{{\rm d}}

\def\cG{{\cal G}}

\def\cL{{\cal L}}

\def\cT{{\cal T}}
\def\cU{{\cal U}}
\def\cV{{\cal V}}

\def\cX{{\cal X}}
\def\cY{{\cal Y}}
\def\cZ{{\cal Z}}

\def\bfx{{\bf x}}
\def\bfy{{\bf y}}

\def\mfg{{\mathfrak g}}

\def\ssA{{\scriptscriptstyle A}}
\def\ssB{{\scriptscriptstyle B}}

\def\ssH{{\scriptscriptstyle H}}
\def\ssI{{\scriptscriptstyle I}}

\def\ssL{{\scriptscriptstyle L}}
\def\ssM{{\scriptscriptstyle M}}
\def\ssN{{\scriptscriptstyle N}}
\def\ssP{{\scriptscriptstyle P}}
\def\ssQ{{\scriptscriptstyle Q}}
\def\ssR{{\scriptscriptstyle R}}

\def\ssT{{\scriptscriptstyle T}}
\def\ssU{{\scriptscriptstyle U}}
\def\ssV{{\scriptscriptstyle V}}

\def\UV{{\scriptscriptstyle U\hbox{\kern-0.1em}V}}
\def\PPN{{\scriptscriptstyle P\hbox{\kern-0.1em}P\hbox{\kern-0.1em}N}}
\def\MN{{\scriptscriptstyle M\hbox{\kern-0.1em}N}}
\def\MNP{{\scriptscriptstyle M\hbox{\kern-0.1em}N\hbox{\kern-0.1em}P}}
\def\KK{{\scriptscriptstyle K\hbox{\kern-0.1em}K}}
\def\SM{{\scriptscriptstyle S\hbox{\kern-0.1em}M}}
\def\EH{{\scriptscriptstyle E\hbox{\kern-0.1em}H}}

\def\QCD{{\scriptscriptstyle Q\hbox{\kern-0.1em}C\hbox{\kern-0.1em}D}}
\def\IR{{\scriptscriptstyle I\hbox{\kern-0.1em}R}}
\def\TEV{{\scriptscriptstyle T\hbox{\kern-0.1em}E\hbox{\kern-0.1em}V}}
\def\aff{{a\hbox{\kern-0.1em}f\hbox{\kern-0.1em}f}}

\title{4D de Sitter from String Theory via 6D Supergravity}
\author{C.P.~Burgess,${}^{1,2,3}$ F.  Muia ${}^4$ and F.~Quevedo${}^{4,5,6}$ \\ 
{\it 
${}^1$ Department of Physics \& Astronomy, McMaster University\\ \qquad 1280 Main Street West, Hamilton ON, Canada.\\
${}^2$ Perimeter Institute for Theoretical Physics\\
\qquad 31 Caroline Street North, Waterloo ON, Canada.\\
${}^3$ School of Theoretical Physics, Dublin Institute for Advanced Studies,\\ \qquad 10 Burlington Rd., Dublin,
  Co. Dublin, Ireland\\
${}^4$ DAMTP, University of Cambridge, Wilberforce Road,  Cambridge, CB3 0WA, UK.\\
${}^5$ CERN, Theoretical Physics Department, 1211, Gen\`eve 23, Switzerland.\\
${}^6$ New York University Abu Dhabi, PO Box 129188, Saadiyat Island, Abu Dhabi, UAE.

}
}

\date{\today}

\abstract{
We propose a new way to obtain explicit de Sitter (dS) solutions from controlled string-theory constructions. The Dine-Seiberg problem is usually interpreted as meaning that weak-coupling expansions generically drive runaways rather than allowing stabilized maximally symmetric spacetimes. Using the special case of string compactifications to 6D we confirm that this argument does prevent the existence of classical maximally symmetric 6D solutions but argue that it {\it allows} time-independent classical solutions with maximal 4D symmetry, including dS solutions. We review how minimal gauged chiral 6D supergravity evades standard dS no-go theorems by having a positive scalar potential and describe the known 4D classical dS, AdS and Minkowski solutions. The stringy provenance of this 6D supergravity was obscure until Grimm and collaborators  found it to be produced by direct F-theory Calabi-Yau flux compactifications. We construct classical 4D maximally symmetric solutions for this 6D supergravity and provide explicit solutions of the higher-dimensional field equations corresponding to dS, AdS and flat spacetimes in 4D, allowing interesting hierarchies of scales. We show how the singularities of these solutions are consistent with the back-reaction of two space-filling 4D brane-like sources situated within the extra dimensions and infer some of the properties of these sources using the formalism of point particle effective field theory (PPEFT). These tools relate the near-source asymptotic forms of bulk fields to source properties and have been extensively tested for more prosaic physical systems involving the back-reaction of small sources, such as the dependence of atomic energy levels on nuclear properties.  We use it to determine the tension of the brane-like sources (that can be positive) and its derivatives. We verify that the solutions are in the weak coupling/large volume regime required to neglect quantum and $\alpha'$ effects.
}

\preprint{CERN-TH-2024-107}

\begin{document}
\maketitle









\section{Introduction}

\begin{quote}
\rightline{{\it The future ain't what it used to be.}}
\end{quote}

\noindent
So far as we know de Sitter space is in our future \cite{Planck:2018vyg}. This is particularly annoying for string theorists because it seems relatively hard to obtain as a solution to that theory's field equations (for a recent review see \cite{Cicoli:2023opf}). So hard that some have conjectured it is a principle of physics that de Sitter space cannot be found in any theory of quantum gravity \cite{Obied:2018sgi}. 

In this paper we find 
explicit solutions to the classical field equations of low-energy string vacua that contain 4D de Sitter space and discuss the challenges to which they point for fully understanding whether or not de Sitter space can be regarded as emerging from a fully fledged string construction.

\subsection{Why it's hard}

Two main obstacles are usually cited when discussing the difficulty of obtaining de Sitter solutions of string theory within a fully controlled (and so trustable) approach.
\begin{enumerate}
\item {\it Classical no-go theorems}. General no-go theorems have been proposed that under very general assumptions exclude the existence of classical 4D de Sitter solutions to the types of equations that typically govern the low-energy effective string theory action \cite{NoGo1, NoGo2, NoGo3}.

\item {\it The Dine-Seiberg problem}. String theory has no parameters and so all approximate expansions involve a series in powers of a field (such as the string dilaton or the inverse of the volume of any extra dimensions). The leading term in the scalar potential is necessarily a monomial in these fields and so drives them to zero or infinity. This suggests that stationary points for these fields should arise only outside the regime of validity of such expansions \cite{Dine:1985he}.
\end{enumerate}

Twenty years of ingenuity have gone into overcoming these challenges and at least two general mechanisms have emerged. In one approach the classical impossibility of de Sitter solutions is accepted but quantum corrections play an important role, with a combination of fluxes, branes, orientifold planes, perturbative and non-perturbative effects cobbled together to generate low-energy effective descriptions whose field equations have 4D de Sitter solutions (see \cite{McAllister:2023vgy} for a recent review and \cite{McAllister:2024lnt} for a recent explicit example). These constructions have many moving parts, however, and debate still rages as to whether all of the corrections are under complete calculational control in the de Sitter regime. Part of the difficulty arises because de Sitter space necessarily breaks supersymmetry and supersymmetry is often crucial for suppressing quantum corrections in explicit constructions. 
 
A second approach instead seeks to evade the classical no-go theorems directly so de Sitter solutions can be found in a regime where all quantum corrections can be systematically neglected. Although conceptually simpler, most attempts so far have failed to find de Sitter solutions completely within a trustable regime (see for instance \cite{Cordova:2018dbb, Andriot:2020wpp}). A key assumption of the no-go theorems is that higher-dimensional scalar potentials are non-positive, as is known to be true for most higher-dimensional supergravities. Most but not all: an exception is the 10D Romans supergravity \cite{Romans} considered in \cite{Cordova:2018dbb}, but its stringy provenance is not fully  understood and so far it is not known how to handle the singularities of the solution. 
 
\subsection{So what's new?}
 
Here we explore a second exception to the negative-potential-in-supergravity rule: gauged chiral 6D supergravity \cite{NS, RDSSS}. This supergravity also has a positive scalar potential and compactified solutions to it have been thoroughly studied for a variety of reasons for over 40 years, starting with the pioneering work of Salam and Sezgin \cite{SS} who found an elegant supersymmetric solution that compactifies to 4D Minkowski space with the extra two dimensions being a sphere whose size is stabilized by electromagnetic flux. 

Further exploration subsequently found a broader class of exact non-supersymmetric classical compactifications \cite{Aghababaie:2003wz, Gibbons:2003di, Aghababaie:2003ar, Burgess:2004dh} to 4D Minkowski space with the extra dimensions deformed into `rugby-ball' geometries with a pair of conical singularities (interpreted as being the positions of a pair of codimension-2 branes). Still later came compactifications to 4D de Sitter and anti-de Sitter \cite{Tolley:2005nu}, again with singularities in the extra dimensions suggesting the presence of source branes. We summarize these solutions in \S\ref{Sec:OldSchool} below. 

It is important that the branes in these solutions are not `probe' branes, since the solutions explicitly include their back-reaction onto the geometry. This back-reaction has two important effects: it causes some bulk fields to become singular at the brane position (much as the Coulomb solution diverges at the position of the source charge) and it can fix some of the remaining moduli associated with the flat directions of the bulk (in principle obviating the need to introduce fluxes to do so). We review a formalism for relating the singular bulk behaviour to source properties in \S\ref{sec:PPEFT}. Ultimately the modulus stabilization occurs because the branes break the approximate symmetries responsible for the moduli \cite{Burgess:2010zt, Burgess:2011mt, Burgess:2013eua}. 

It is noteworthy that this occurs without running afoul of the Dine-Seiberg problem, and does so in an interesting way. The 6D scalar potential indeed arises as expected as a monomial in the 6D expansion field (a dilaton) in precisely the expected way, and the rolling of the dilaton that this implies indeed obstructs the existence of a maximally symmetric {\it six-dimensional} solution. It does {\it not} however obstruct the existence of solutions that are maximally symmetric only in four dimensions (as is 4D de Sitter space) rather than in the full six dimensions. In the solutions we find the dilaton evolves in space and not time and does so in a way that extrapolates between the boundary conditions imposed by the presence of source branes. In this way the Dine-Seiberg problem gets recast from a bug into a feature: it helps explain why static solutions compactify spacetime (see \cite{Burgess:2008az} for a related discussion). 
 
A perceived problem with these solutions from the string point of view has been the absence of a clear stringy pedigree for the 6D supergravity. At present there are two separate responses to this objection:
\begin{itemize}
\item We briefly review the arguments of \cite{Cvetic:2003xr} stating that {\it any} solution to the 6D chiral supergravity of interest uplifts to a solution of the full ten-dimensional field equations of Type-I and heterotic supergravites. This argument uses an explicit consistent truncation on a noncompact extra dimension with infinite volume -- and so does not provide a finite prediction for the 6D Newton constant (or a viable low-energy phenomenology). But the ability to uplift to ten dimensions known 6D compactifications with 4D de Sitter space arguably can be regarded as an existence proof for 4D de Sitter solutions to a class of 10D field equations of a supergravity with a known stringy provenance.
\item To show why the issue of noncompact extra dimensions is ultimately a red herring we work in this paper with a closely related 6D chiral supergravity that was found to be a low-energy limit for $F$-theory compactifications in Type IIB string vacua with compact Calabi-Yau-like extra dimensions \cite{Grimm:2013fua}. The 6D theory found in this way differs by the addition of several fields relative to the 6D theory explored in {\it e.g.}~\cite{Tolley:2005nu} and so in this paper we repeat the exercise of constructing 4D de Sitter solutions to the theory that results, thereby providing a new class of explicit 4D de Sitter compactifications as solutions to 10 equations with a bona-fide stringy pedigree (but this time with only compact extra dimensions).  
\end{itemize}
These two types of uplifts from 6D to 10D are reviewed in \S\ref{Sec:10DUplift} and the new class of compactifications from $F$-theory are described in \S\ref{Sec:NewWave}.  

So does this settle the issue of de Sitter solutions existing in string theory? Not quite, though it does move the ball downfield somewhat. A remaining concern involves the singularities these solutions have, which raise two separate issues. The first asks whether any small expansion fields become large near the singularities, since this can threaten the validity of the approximations made when asserting the solutions capture the properties of real string vacua. The second assumes the singularities can be interpreted as indicating the presence of some sort of brane source and asks what can be learned about the source properties. In particular, are these recognizable as elementary objects like D-branes or orientifold planes that are known to arise within string theory? 

We do not yet have definitive answers to these questions, but experience elsewhere in physics teaches us that neither issue need be fatal in itself. Indeed the Coulomb solution remains beloved despite being famously singular at the origin. Furthermore this singularity is not in itself an obstacle to controlled calculations of atomic properties. Indeed, this atomic analogy is informative in several ways since we there understand what is going on in great detail. The singularity at $r=0$ indicates the existence of a localized object (the nucleus) that sources the field, and the singularity only arises when the idealized external Coulomb potential is naively extrapolated into the interior of the nucleus instead of using the field the actual nucleus really generates. 

We also know that strong couplings emerge at short distances that bind the nuclei and cannot be captured purely using electromagnetic reasoning. Yet for the purposes of computing atomic electronic energy levels all of the uncertainties associated with these are well-described by an effective theory that systematically captures how localized first-quantized sources interact with low-energy electromagnetic and electron fields in their immediate surroundings (for a review see \cite{EFTBook}). The use of this effective theory does not undermine in any way our ability to accurately compute atomic energy levels, including the leading contributions from nuclear structure.  

The formalism -- point-particle effective field theory, or PPEFT -- for systematically determining how individual (first-quantized) small compact objects affect their surroundings was developed \cite{Burgess:2016lal, Burgess:2016ddi, Burgess:2017ekj} and tested in some detail for practical systems like nuclei in atoms \cite{Burgess:2017mhz, Plestid:2018qbf, Zalavari:2020tez, Burgess:2020ndx}. It is also known to capture the correct matching between source and environment -- such as between D-branes and their surroundings (see {\it e.g.}~\cite{Bayntun:2009im}) -- for gravitational systems \cite{Burgess:2008yx}. This effective theory provides a connection between the asymptotic near-source form of any bulk fields and the dependence of the source's action on these fields, in much the same way that Gauss' Law provides a relation between the coefficient of $1/r$ in the Coulomb field and a source's electric charge within electromagnetism. 

We apply this formalism to the singularities of the 4D de Sitter solutions and find a relationship between the near-source asymptotic form of these solutions and the effective PPEFT action describing these sources. This connection does not completely determine the microscopic form of the source any more than measurements of electronic energy levels can completely determine the quark structure of the atomic nucleus. But they do efficiently characterize precisely what any brane sources must satisfy in order to source 4D de Sitter geometries within the supergravities of interest. This leaves open whether more microscopic constructions using known stringy constituents can be found that have the required properties.      

We conclude with a brief summary of our results in \S\ref{Sec:Conclusions}.
 
\section{4D de Sitter from 6D supergravity }
\label{Sec:OldSchool}

\begin{quote}
\rightline{{\it If you don't know where you are going you might wind up someplace else.}} 
\end{quote}

\noindent
This section reviews chiral 6D supergravity and summarizes its 4D solutions (including those with nonzero 4D curvature) and what is known about their singularities and what these say about the gravitating sources that generate these solutions. 

\subsection{Chiral 6D Supergravity}

Chiral gauged 6D $(1,0)$ supergravity \cite{NS, RDSSS} contains a single chiral supersymmetry in six dimensions. The field content is built from the basic supersymmetric multiplets, which include:

\begin{itemize}
\item {\it Gravity multiplet:} Metric $g_{\ssM\ssN}$, self-dual antisymmetric tensor $B_{\ssM\ssN}^+$, left-handed gravitino $\Psi_\ssM^\alpha$ for a total of 12 bosonic and 12 fermionic degrees of freedom.
\item{\it Tensor multiplet.} Anti-self-dual skew tensor $ B_{\ssM\ssN}^-$, scalar $\varphi$, right-handed fermion $\psi$ (tensorino) for a total of 4 bosonic and 4 fermionic degrees of freedom.
\item{\it Vector multiplet.} Gauge potential $A_\ssM$ and right-handed fermion $\lambda$ (gaugino) for a total of 4 bosonic and 4 fermionic degrees of freedom.
\item{\it Hypermultiplet:} Two complex scalars $q^1, q^2$ and one right-handed Weyl fermion $\xi$ (hyperino) for a total of 4 bosonic and 4 fermionic degrees of freedom.
\end{itemize}

In general there can be more than one of each type (except the graviton multiplet), with $n_\ssT$ denoting the number of tensor multiplets, $n_\ssV$ the number of vector multiplets and $n_\ssH$ the number of hypermultiplets. Indeed, anomaly cancellation implies multiple multiplets are the rule not the exception. Although having chiral fermions is attractive for phenomenological purposes it also means that care must be taken to ensure that no gauge symmetries are anomalous. Green-Schwarz anomaly cancellation \cite{GS} can occur in 6D (just as it does in 10D) but only if some consistency conditions are satisfied, such as the number of each type of multiplet satisfies  \cite{Green:1984bx, Sagnotti:1992qw,Erler:1993zy}
\begin{equation}
n_\ssH - n_\ssV + 29 n_\ssT =273 \,.
\end{equation}
The number of scalar fields depends on the number of tensor and hypermultiplets, with the ones from the tensor multiplets parametrizing the coset $SO(1,n_\ssT)/SO(n_\ssT)$ while the $4 n_\ssH$ scalars coming from the hypermultiplets parametrize a quaternionic manifold\footnote{See \cite{Hamada:2023zol,Becker:2023zyb,Loges:2024vpz,Hamada:2024oap,Baume:2024oqn,Kim:2024tdh} for recent discussions related to a general classification of chiral supersymmetric 6D theories.}. 

6D chiral theories are notorious because for them an action need not always exist if there are unequal numbers of self-dual and anti-self-dual skew tensor fields. For this reason and for simplicity we concentrate on the case $n_\ssT=1$ so that the field $B_{\ssM\ssN}^-$ from the tensor multiplet can combine with $B_{\ssM\ssN}^+$ from the gravity multiplet into an unconstrained antisymmetric tensor $B_{\ssM\ssN}$. In this case we have a single tensor-multiplet scalar $\varphi$ and an action formulation exists. The single tensor-multiplet scalar we denote by\footnote{In general for $n_\ssT > 1$ the coset  $SO(1,n_\ssT)/SO(n_\ssT)$ is parametrized by fields $j^\alpha$, with $\alpha=1, \cdots, n_\ssT+1$, constrained by $\Omega_{\alpha\beta}j^\alpha j^\beta=1$, with $\Omega_{\alpha\beta}=\text{diag}(-1,1,\cdots, 1)$. For $n_\ssT=1$, the constrain is satisfied by $j^0=\cosh\varphi$, $j^1=\sinh \varphi$ and we can work with the unconstrained field $\varphi$. See for instance \cite{Ferrara:1997gh}.} $\varphi$.

When searching for background configurations we can set all fermion fields to zero and focus on scalar fields, that come from the tensor multiplets and hypermultiplets. Because the supergravity is gauged there is a scalar potential of the form \cite{NS}
\be
  V = \frac{2\mfg^2}{\kappa^4} \, U(q) \, e^{\varphi} \,,
\ee
where $\mfg$ is the coupling for a specific 6D gauge field and $\kappa^2 = 8\pi G_6$ is the 6D Newton's constant for gravity. The function $U(q)$ can be minimized for the hypermultiplet fields, $q^\ssU$, leading to a vacuum configuration for which these fields can all be set to zero consistent with their equations of motion.\footnote{The detailed form of the function $U(q)$ depends on which groups are gauged but it takes generically the small-$q$ form $U(q)= A+\sum_i B_i |q_i|^2$ with $A,B_i$ positive constants, where the sum is over a subset of the hypermultiplets \cite{Nishino:1986dc,Avramis:2005qt,Randjbar-Daemi:2004bjl,Suzuki:2005vu}.} The function $U(q)$ is normalized so that $U_{\rm min} = 1$ at this minimum. (In later sections we return to allow nontrivial hypermultiplet scalar configurations when we discuss $F$-theory.) 

Background gauge fields can be nonzero consistent with maximal symmetry in 4D and we consider only a single nonzero background gauge field, chosen to be the gauge potential, $A_\ssM$, that gauges the specific $U_\ssR(1)$ symmetry for which the gravitino field carries nonzero charge. The action for this gauge field, the Kalb-Ramond field $B_{\ssM\ssN}$ the remaining scalar $\varphi$ and the metric then has the form considered in the original paper of Salam and Sezgin:\footnote{Like all sensible people we use a positive-signature metric and Weinberg's curvature conventions \cite{GnC}, which differ from the popular MTW conventions \cite{MTW} only in the overall sign of the Riemann tensor.}
\be   \label{salamsezgin}
    \cL_6 = - \sqrt{-g} \left[ \frac{1}{2 \kappa^2} \,
    g^{\ssM\ssN} \Bigl( R_{\ssM\ssN} + \partial_\ssM \varphi \,
    \partial_\ssN \varphi \Bigr) + \frac{1}{4}  e^{-  \varphi} F_{\ssM\ssN} F^{\ssM\ssN}+ \frac{1}{12}  e^{-  2\varphi} H_{\ssM\ssN\ssP} H^{\ssM\ssN\ssP}
    + \frac{2\mfg^2}{\kappa^4} \, e^{\varphi} \right] \,,
\ee
%
with field strengths defined by $F_{(2)} = \exd A_{(1)}$, $H_{(3)}=\exd B_{(2)}+\frac12 \kappa F_{(2)}\wedge A_{(1)}$ (where the bracketed subscript $(p)$ in the index-free notation indicates the field in question is a $p$-form).

The bosonic equations of motion are:
\bea \label{SSFEvarphi}
    &&\qquad\Box_6 \varphi +\frac{\kappa^2}{4} e^{-\varphi} F_{\ssM\ssN}F^{\ssM\ssN}+\frac{\kappa^2}{6} e^{-2\varphi} H_{\ssM\ssN\ssP}H^{\ssM\ssN\ssP}-
    \frac{2{\mfg}^2}{\kappa^2}e^{\varphi} =0 \\
   \label{SSFForm}
    &&\nabla_\ssM \Bigl( e^{-\varphi} F^{\ssM\ssN} \Bigr)+ \kappa e^{-2\varphi} H^{\ssP\ssN\ssQ} F_{\ssP\ssQ} =0 \,,\qquad\qquad \nabla_\ssM \Bigl( e^{-2\varphi} H^{\ssM\ssN\ssP} \Bigr) =0  \\
    \label{SSFEinstein}
    &&\qquad\qquad R_{\ssM\ssN} + \partial_\ssM \varphi \, \partial_\ssN \varphi+\kappa^2
    e^{-\varphi}F_{\ssM\ssP}F_{\ssN}^\ssP+\frac{1}{2}(\Box_6 \varphi) \,g_{\ssM\ssN} =0 \,, 
\eea
in which \pref{SSFEvarphi} has been used to rewrite \pref{SSFEinstein} so that the terms proportional to $g_{\ssM\ssN}$ involve only the 6D d'Alembertian $\Box_6 \varphi := g^{\ssP\ssQ} \nabla_\ssP \nabla_\ssQ \varphi$. This turns out to be possible because the system has a classical scaling symmetry under which the replacements
\be \label{SSRescaling}
g_{\ssM\ssN} \rightarrow c\, g_{\ssM\ssN} \quad \hbox{and} \quad e^{-\varphi} \rightarrow c \, e^{-\varphi}\qquad \hbox{imply} \qquad  {\mathcal L}_6\rightarrow c^2 {\mathcal L}_6 \,,
\ee
for constant $c$. Although not a symmetry of the action this transformation does leave the equations of motion invariant \cite{Aghababaie:2003ar, Burgess:2011rv, Gautason:2013zw}.

\subsection{Solutions}

Because the scalar potential for $\varphi$ is monotonic it obstructs there being 6D maximally symmetric solutions for any finite $\varphi$. This can be seen because maximal symmetry in 6D requires $F_{\ssM\ssN} = H_{\ssM\ssN\ssP} = 0$ and that $\varphi$ be constant, but this is inconsistent with the dilaton field equation which implies $\Box_6 \varphi$ cannot vanish. The same is not true for solutions that are maximally symmetric only in 4D, however, since although these still require $H_{\ssM\ssN\ssP}=0$ the gauge field can be nonzero if restricted to the two extra dimensions: $F_{mn}$  with $m,n=4,5$. 

\subsubsection{Salam-Sezgin solution}

For the simplest solutions spacetime has a product metric, 
\be 
  \exd s^2=g_{\mu\nu}(x) \, \exd x^\mu \exd x^\nu+g_{mn}(y) \, \, \exd y^m \exd y^n,
\ee
with $g_{mn}$ being the metric on a 2-sphere and $F_{mn}= f \, \epsilon_{mn}$ proportional to the 2-sphere volume form. Maximal 4D symmetry requires $\varphi$ and $f$ are independent of the 4D coordinates $x^\mu$ and the field equation \pref{SSFForm} then implies $e^{-\varphi} f$ is also independent of the coordinates $y^n$. For $f$ and $\varphi$ separately constant $\Box_6 \varphi = \Box_2\varphi = 0$ and consistency with \pref{SSFEvarphi} requires 
\be \label{varphifval}
   e^{-\varphi} f = \pm 2\mfg/\kappa^2.
\ee 
 
When $\Box_6 \varphi = 0$ eq.~\pref{SSFEinstein} also implies the maximally symmetric 4D metric $g_{\mu\nu}$ must be flat. The resulting maximally symmetric 4D and 2D geometry then is ${\mathbb R}^{1,3}\times S^2$ with ${\mathbb R}^{1,3}$, with metric
\be
 \exd s^2=\eta_{\mu\nu}\, \exd x^\mu \exd x^\nu+\rho^2\left(\exd\phi^2+\sin^2\phi \,\exd \theta^2\right)
\ee
where $\rho$ is the 2-sphere's radius. The radius is determined by the extra-dimensional components of \pref{SSFEinstein} -- $R_{mn}=- \kappa^2 e^{-\varphi}F_{mp}F^p_n$ -- which imply
\be\label{minimum}
  \frac{1}{\rho^2}=\kappa^2 f^2e^{-\varphi}  \quad \hbox{which with \pref{varphifval} implies} \quad
   \rho^2e^{\varphi}=\left(\frac{\kappa}{2\mfg}\right)^2 \,.
\ee
Flux quantization on the 2-sphere also implies $\tilde \mfg f \rho^{2} = \frac12 n$ for integer $n$, where $\tilde \mfg$ is the gauge coupling for the field $F_{\ssM\ssN}$. This is consistent with \pref{varphifval} and \pref{minimum} if $\tilde \mfg = \mfg$ and $n = \pm 1$. This choice is called the Salam-Sezgin solution.

Notice that the field equations determine only the combination $\rho^2 e^{\varphi}$, with the quantities $\rho$ and $\varphi$ not separately determined. This is consistent with the classical rescaling symmetry \pref{SSRescaling}, which implies one combination of these two moduli is a flat direction of the potential in the 4D effective theory. 

The Salam-Sezgin solution is remarkable in several ways. It happens that the choice $n = \pm 1$ preserves ${\mathcal N}=1$ supersymmetry in 4D, so these compactifications provide explicit chiral ${\mathcal N}=1$ Minkowski solutions -- much as Calabi-Yau compactifications do for string theory -- but in a simpler framework for which the metric is known explicitly. Furthermore, since the starting 6D theory has a positive runaway potential, it illustrates directly the Dine-Seiberg problem, not allowing a maximally symmetric solution in 6D. But the possibility of turning on magnetic  fluxes allows a maximally symmetric solution in 4D in which the fluxes compete with the runaway scalar potential to give rise to a 4D Minkowski vacuum.

From the 4D point of view the 4D scalar potential for the fields $\varphi$ and $\rho$ receives contributions at the classical level from: (i) the 6D scalar potential, (ii) the contribution of 2D curvature to the Einstein-Hilbert term of the action and (iii) the contribution of magnetic flux to the Maxwell action. It happens that for $n = \pm 1$ these combine to a perfect square, leading to the 4D scalar potential \cite{Aghababaie:2002be}
\be
V=\frac{2\mfg^2e^\varphi}{\rho^2}\left(1-\frac{\kappa^4}{4\mfg^2 e^\varphi \rho^2}  \right)^2 \,.
\ee
This reproduces the solution in \pref{minimum} inasmuch as it fixes the combination $e^\varphi \rho^2$ but leaves the orthogonal combination $e^\varphi/\rho^2$ unfixed.\footnote{In \cite{Aghababaie:2002be} non-perturbative effects were added to fix the remaining flat direction. The remaining flat direction arises because the Salam-Sezgin solution does not break the scaling symmetry \pref{SSRescaling}.}

The Salam-Sezgin solution is known to be a particular instance of a broader class of solutions to these equations that include both warped geometries and conical singularities in the extra dimensions \cite{Gibbons:2003di, Aghababaie:2003ar, Burgess:2004dh} (for all of which the 4D metric $g_{\mu\nu}$ remains flat). 

\subsubsection{A broader class of solutions}

A wide variety of exact solutions to the 6D equations \pref{SSFEvarphi} through \pref{SSFEinstein} are known, including scaling solutions \cite{Tolley:2006ht}, wave solutions \cite{Tolley:2007et}, black brane solutions \cite{Burgess:2014qha} and so on. Of particular interest here are those described in \cite{Tolley:2005nu} for which the 4D geometry is maximally symmetric but not flat.

These solutions are obtained starting with $H_{\ssM\ssN\ssP} = 0$ and seeking geometries with the warped-product metric {\it ansatz}
\be \label{6DMetricAnsatz}
    \exd s^2= \hat{g}_{\ssM\ssN} \, \exd x^\ssM \exd x^\ssN =  W^2 g_{\mu\nu} \,\exd x^{\mu} \exd x^{\nu}  +a^2 \exd\theta^2+a^2W^8 \exd\eta^2 \,,
\ee
where $g_{\mu\nu}(x)$ is a maximally symmetric 4D metric and the `hat' is meant to distinguish $\hat g_{\mu\nu} = W^2 g_{\mu\nu}$ from $g_{\mu\nu}$. The unknown functions $W(\eta)$, $a(\eta)$, $\varphi(\eta)$ and $F_{\eta\theta}(\eta)$ are defined on the cylindrically symmetric extra-dimensional 2D geometry spanned by the coordinates $\theta$ and $\eta$. 

Notice that maximal 4D symmetry implies the 4D components of \pref{SSFEinstein} become
\be \label{R4isBox}
  \hat R_{\mu\nu} +\frac{1}{2}(\Box_2 \varphi) \, \hat g_{\mu\nu} =0 \,,
\ee
where $\hat R_{\mu\nu} = \frac14 \,\hat R_4 \,\hat g_{\mu\nu}$. This shows that $\hat R_4 = \hat g^{\mu\nu} \hat R_{\mu\nu} = - 2 \Box_2 \varphi$ is a total derivative within the two extra dimensions. If fields are everywhere nonsingular then $\oint \Box_2 \varphi = 0$ when integrated over the extra dimensions and so the maximally symmetric 4D geometry must be flat. But singularities are not unusual when gravitating sources are present (as the Coulomb solution teaches us for electromagnetic sources), so a better approach is to excise the positions of any singular sources by surrounding each of them by a small Gaussian pillbox. In this case integrating over the extra dimensions outside the pillboxes picks up contributions from the pillbox surfaces, and provides a constraint relating the sign of $\hat R_4$ to the near-source derivative of $\varphi$: 
\be \label{R4vsSurfTerms}
    \int_{M_2} \exd^2 y \, \sqrt{g_2} \, \hat R_4(x,y) = 2 \oint_{\partial M_2} \exd y \; n_m \sqrt{g_2} \, g^{mn} \partial_n \varphi \,,
\ee
where $\sqrt{g_2}$ is the 2D volume element and $\partial M_2$ is the union of the surfaces of the small pillboxes, on which $n_m$ is the unit normal (pointing out of the source) \cite{Aghababaie:2003ar, Burgess:2011rv, Gautason:2013zw}.  

With these choices the Maxwell equation \pref{SSFForm} integrates to give
\be \label{FetathetaSoln}
   F_{\eta \theta} = Q \, a^2  \, e^{\varphi} \,,
\ee
for constant $Q$ while the functions $a$, $W$ and $\varphi$ are obtained by integrating the scalar equation \pref{SSFEvarphi}
\be \label{dilatoneqn}
    \varphi'' + \frac{\kappa^2}{2}Q^2a^2
    e^{\varphi} -\frac{2 {\mfg}^2}{\kappa^2}a^2W^8e^{\varphi}
    =0 \,,
\ee
and the Einstein equations \pref{SSFEinstein}
\bea \label{einsteineqn1}
    &&(\mu\nu): \quad
    \left(   \ln W + \frac{\varphi}2 \right)''
    = 3 \zeta\, H^2a^2 W^6\\
    \label{einsteineqn2}
    &&(\theta\theta): \quad
    \left( \ln a + \frac{\varphi}{2} \right)'' = - \kappa^2 Q^2 \, a^2 e^\varphi
\eea
where primes denote differentiation with respect to $\eta$. Here $\zeta = \pm 1$ and $H$ is the maximally symmetric 4D geometry's Hubble constant, in terms of which the curvature of the 4D metric $g_{\mu\nu}$ is $R_{\mu\nu\lambda\rho} = \zeta  H^2 (g_{\mu\rho} g_{\nu\lambda} - g_{\mu\lambda} g_{\nu\rho})$ (and so $\zeta = +1$ for de Sitter space and $\zeta = -1$ for anti-de Sitter space). 

Eqs.~\pref{dilatoneqn} through \pref{einsteineqn2} are to be read as evolution equations for stepping the three unknown functions $a$, $W$ and $\varphi$ forward in $\eta$ given initial conditions at some $\eta = \eta_0$. These initial conditions must satisfy the first-order constraint coming from the $(\eta\eta)$ Einstein equation,
\be \label{einsteinconstraint}
    (\eta\eta): \quad 6\zeta H^2 a^2 W^6 - \frac{4\,a'
    W'}{aW} - \frac{6(W')^2}{W^2} + \frac12 \, (\varphi')^2 +
    \frac{\kappa^2}{2} Q^2 \, a^2 e^\varphi - \frac{
    2{\mfg}^2}{\kappa^2} \, a^2 W^8 e^\varphi = 0 \,,
\ee
from which the value of $\zeta H^2$ can be read off once the fields and their first derivatives are specified for some initial value of $\eta$. The scale invariance \pref{SSRescaling} of the full 6D field equations implies a one-parameter family of solutions can be built from any specific solution, with 
\be \label{SSdegeneracy}
 \Bigl\{\varphi , a, W, H \Bigr\} \to \Bigl\{ \varphi + \varphi_0 ,
 a\, e^{- \varphi_0/2} , W, H \, e^{\varphi_0/2}  \Bigr\} \,,
\ee
for $\varphi_0$ an arbitrary real constant. For the special case $H = 0$ this corresponds to a one-parameter family of classical solutions all sharing the same Hubble rate, and so corresponding to a flat direction (labelled by $\varphi_0$) of the 4D effective theory that represents a compactification modulus.

Notice that \pref{einsteineqn2} implies $(\ln a + \tfrac12 \varphi)'$ is a monotonically decreasing function of $\eta$ while \pref{einsteineqn1} implies $(\ln W + \tfrac12 \varphi)'$ is monotonically increasing or decreasing depending on the sign of $\zeta$. Integrating \pref{einsteineqn1} specializes \pref{R4vsSurfTerms} to this geometry  
\be  \label{IntH2}
   3 \zeta\, H^2 \int_{\eta_1}^{\eta_2} \exd \eta \; a^2 W^6 = \left( \ln W + \frac{\varphi}{2} \right)'_{\eta=\eta_2} -  \left( \ln W + \frac{\varphi}{2} \right)'_{\eta = \eta_1} \,.
\ee 
This last equation has a simple 4D interpretation if integrated over the entire extra dimensions once the definition 
\be \label{4DkappaDef}
  \frac{1}{\kappa^2_4} = \frac{1}{\kappa^2} \int \exd^2x  \; \sqrt{g_2} \; W^2 =  \frac{2\pi}{\kappa^2} \int_{-\infty}^\infty \exd \eta \; a^2  W^6 
\ee
of the low-energy 4D Newton constant, $\kappa_4^2 = 8 \pi G_\ssN$, is used because together \pref{IntH2} and \pref{4DkappaDef} imply
\be \label{4DFriedmann}
    \zeta H^2 = \tfrac83 \pi G_\ssN \left[ \left(\,\ln W + \frac{\varphi}{2} \right)'\,
    \right]_{\eta = -\infty}^{\eta = + \infty}\,.
\ee
This is recognizable as the 4D Friedmann equation, with the role of the 4D energy density being played by the square bracket involving only the asymptotic derivatives of the combination $W + \tfrac12 \varphi$. This expression can be shown to be completely consistent with the 4D Friedmann equation found within the low-energy 4D EFT once its effective scalar potential is carefully computed \cite{Bayntun:2009im, Burgess:2010zt, Burgess:2011mt, Burgess:2015nka, Burgess:2015lda}.

Explicit solutions to these equations with $H = 0$ are known in closed form \cite{Gibbons:2003di, Aghababaie:2003ar, Burgess:2004dh}
\bea \label{Flat6DSolutions}
  &&W^2 e^\varphi = e^{-\lambda_3 \eta} \,, \quad
  W^4 = \left( \frac{\kappa^2 Q\lambda_2}{2\mfg\lambda_1} \right) \frac{\cosh[\lambda_1(\eta - \eta_1)]}{\cosh[\lambda_2(\eta - \eta_2)]} \,,
  \quad F_{\eta\theta} = \left( \frac{Q a^2}{W^2} \right) e^{-\lambda_3 \eta} \nn\\
  &&\qquad\hbox{and} \qquad 
  \frac{1}{a^4} = \left( \frac{4\mfg\kappa^4Q^3}{\lambda_1^3 \lambda_2} \right) e^{-2\lambda_3\eta} \cosh^3[\lambda_1(\eta-\eta_1)] \cosh[\lambda_2(\eta - \eta_2)]   \,,
\eea
with integration constants $\eta_1$, $\eta_2$, $\lambda_1$, $\lambda_2$ and $\lambda_3$ related by $\lambda_2 = \sqrt{\lambda_1^2 + \lambda_3^2}$. There is no loss of generality in choosing $\lambda_2 \ge 0$, in which case it must satisfy $\lambda_2 \ge |\lambda_1|$ (with equality if and only if $\lambda_3 = 0$). A one-parameter family of solutions can be obtained by acting on this with the transformation \pref{SSdegeneracy}. The stability of these solutions is extensively explored in \cite{Burgess:2006ds}.

Solutions also exist for either sign of $\zeta$ and for nonzero $H$, though not analytically in closed form as above. Several explicit examples with 4D de Sitter space ($\zeta = +1$ and $H \neq 0$) are obtained numerically in \cite{Tolley:2005nu}. 

\subsubsection{Near-brane asymptotics}

The solutions \pref{Flat6DSolutions} are generically singular as $\eta \to \pm \infty$ (as are also the solutions with $H \neq 0$), with a curvature singularity when $\lambda_3 \neq 0$ and a conical singularity when $\lambda_3 = 0$. The $\lambda_3 = 0$ solutions include  (but are not restricted to) the unwarped, constant-dilaton `rugby ball' configurations of ref.~\cite{Aghababaie:2003wz} as the special case where $\eta_1 = \eta_2$. These singularities are interpreted as signalling the presence of some sort of a gravitating source, and so we explore the near-source asymptotic form following \cite{Bayntun:2009im}.

If the source is a codimension-two object then we must ask $a \to 0$ as it is approached (so that circles of proper radius $r$ that surround it also have circumferences that shrink as $r \to 0$). In the $a \to 0$ limit eqs.~\pref{dilatoneqn}, \pref{einsteineqn1} and \pref{einsteineqn2} simplify to
\be
 \varphi'' \simeq \left( \ln W \right)'' \simeq \left( \ln a \right)'' \simeq 0 \,,
\ee
and so for the sources at $\eta=\pm\infty$ we have (respectively)
\be \label{asymptoticpowers}
 \varphi \simeq \mp  \eta \, q_\pm  \,, \quad
 W \simeq W_\pm \,  e^{\mp \eta \, \omega_\pm }
 \quad \hbox{and} \quad  a \simeq a_\pm\, e^{\mp \eta \, \alpha_\pm} \,,
\ee
for independent real constants $\alpha_b$, $\omega_b$ and $q_b$ applying for the two limits, $\eta \to \pm \infty$. The explicit $\mp$ sign is present so that the convention is that the functions $W$, $a$ and $e^{\varphi}$ all tend to zero at the position of the source if the constants $\alpha_\pm$, $\omega_\pm$ and $q_\pm$ are positive (and so asking $a \to 0$ at each brane is equivalent to requiring $\alpha_\pm > 0$).  The constraint eq.~\pref{einsteinconstraint} implies the relation 
\be \label{qconstraint}
 q_b^2 = 4\omega_b (2 \alpha_b + 3 \omega_b) \,,
\ee
must hold separately for both choices $b = \pm$.

Even if $a \to 0$ at the source, neglect of the quantities $a^2 W^6$, $a^2 e^\varphi$ and $a^2 W^8 e^\varphi$ in eqs.~\pref{dilatoneqn}
through \pref{einsteinconstraint} is only consistent if $W$ and $e^\varphi$ do not grow fast enough to overwhelm the shrinking of $a$, which requires
\be \label{inequalities}
 2\alpha_b + 6 \omega_b > 0 \,, \quad
 2\alpha_b + q_b > 0 \quad \hbox{and} \quad
 2\alpha_b + 8 \omega_b + q_b > 0 \,.
\ee
The first of these also ensures convergence of the integral appearing in \pref{4DkappaDef}. Notice also that if $\alpha_b > 0$, then $\omega_b$ must also be non-negative. To see this, suppose $\omega_b$ were negative. Then eq.~\pref{qconstraint} would imply $-2\alpha_b - 3\omega_b > 0$, and so adding this to the first of eqs.~\pref{inequalities} would give $\omega_b > 0$ (contradicting the assumption that it is negative). In principle, the constant $q_b$ can have either sign.

When the inequalities \pref{inequalities} hold the terms involving $H$ in the equations of motion become negligible in the near-brane limit. This means that the asymptotic form of the explicit $H=0$ solutions of \pref{Flat6DSolutions} are also relevant to the asymptotic regions at $\eta \to \pm \infty$ when $H \neq 0$. In particular, the metric singularities mentioned above for the flat case also apply. Comparing \pref{asymptoticpowers} with the limiting forms of \pref{Flat6DSolutions} and using the above-mentioned condition $\lambda_2 \ge |\lambda_1|$ shows that  
\be \label{GGPpowers}
 \alpha_\pm = \frac14 \left(3\lambda_1 + \lambda_2 \mp  2 \lambda_3 \right)\ge 0\,, \quad
 \omega_\pm = \frac14 \left( \lambda_2 - \lambda_1 \right) \ge 0\,,
\ee
and
\be
 q_\pm = \pm \lambda_3 - \frac12 \left( \lambda_2 - \lambda_1 \right) \,.
\ee

The quantity $(\ln W + \tfrac12 \varphi)' \simeq \omega_b + \tfrac12 q_b$ therefore asymptotes to 
\be \label{omegaGGP}
 \omega_\pm + \frac{q_\pm}{2} = \pm \frac{\lambda_3}{2} \,,
\ee
and so \pref{4DFriedmann} governing the size of $H$ becomes
\be \label{IntegratedCBA-GGP}
    \zeta H^2 
    = - \tfrac83 \pi G_\ssN  \sum_{b=\pm} \left(\frac{q_b}{2} + \omega_b  \right) \,.
\ee
Although this vanishes for the solutions \pref{Flat6DSolutions} (as it must) once \pref{omegaGGP} is used, the result need not be zero for more general solutions because eqs.~\pref{GGPpowers} need hold only approximately in the near-brane region, with constants $\lambda_1$, $\lambda_2$ and $\lambda_3$ that generically are not the same for the asymptotic region near different sources. 

Because $\omega_b$ is non-negative eq.~\pref{IntegratedCBA-GGP} implies a 4D de Sitter solution (for which $\zeta > 0$) requires at least one of the $q_b$'s to be negative (and so $e^\varphi$ would diverge if $|\eta|$ were really taken to infinity for this source). Such divergences are sometimes dealt with by `smearing' them, 
as one might do if the source of a Coulomb potential were at close examination found actually to be a classical distribution of charge rather than a point source. But as the example of the nucleus within an atom shows, what is really going on at short distances can be much more complicated than a classical distribution of charge, so smearing can at best be regarded as an {\it ad hoc} model of the real short-distance physics. Much better to be more systematic: use EFT methods to characterize the kinds of fields that can emerge from any particular Gaussian pillbox.  


\section{Effective Field Theory of Localized Sources}
\label{sec:PPEFT}

\begin{quote}
\rightline{{\it In theory there is no difference between theory and practice. In practice there is.}}
\end{quote}

\noindent
The near-source singular behaviour found above raises two related questions, about both of which EFT methods have something to say:
\begin{itemize}
\item Do near-source singularities imply an irretrievable breakdown of the approximations that allow reliable predictions to be made for properties (like the 4D curvature) far from the sources? 
\item What does singular behaviour say about the nature of the gravitating sources whose presence is responsible for the solution's singularity?  
\end{itemize}
Effective field theories (EFTs) shed light on both of these questions by systematizing the types of couplings that can arise between small massive sources and surrounding `bulk' fields, organizing them in powers of small ratios of scale like $R/\lambda$ where $R$ is the source's size and $\lambda$ is the wavelength of modes of the bulk field.

Several flavours of such EFTs are fairly widely used. Theories like NRQED \cite{Caswell:1985ui} -- applied to bound states in QED -- or HQET \cite{HQET} -- applied to heavy-quark systems in QCD -- formulate both the bulk fields and the heavy sources as second-quantized fields. For instance in this framework the interactions between the electromagnetic field $A_\mu(x)$, the electron field, $\Psi(x)$, (which charge $-e$ and mass $m$) and the field $\Phi(x)$ for a spin-half nucleus with charge $Ze$ and mass $M\gg m$ can be written as $S= S_{Ren}+S_{NRen}$ with the renormalizable action describing the standard minimal couplings:
 \be
 S_{Ren}=S_{QED}+S_\Phi=-\int \exd^4x\; \left\{\frac{1}{4}F_{\mu\nu} F^{\mu\nu}+\overline\Psi\left[\cancel{D}+m\right]\Psi+\overline\Phi\left[\cancel{D}+M\right]\Phi\right\}.
 \ee
Here $F_{\mu\nu}=\partial_\mu A_\nu-\partial_\nu A_\mu$, $D_\mu\Psi=\partial_\mu\Psi +ieA_\mu\Psi$ and $D_\mu\Phi=\partial_\mu\Phi-iZeA_\mu\Phi$. 

The non-renormalizable action captures the non-minimal interactions among the fields:
\be
S_{NRen}=-\int \exd^4x\; \left\{\frac{\tilde c_d}{2}\left(\overline\Phi \gamma^{\mu\nu}\Phi\right) F_{\mu\nu}+\tilde c_s\left(\overline\Psi\Psi\right)\;\left(\overline\Phi\Phi\right)+\tilde c_v \left(\overline\Psi\gamma^\mu\Psi\right)\left(\overline\Phi\gamma_\mu\Phi\right)+\cdots \right\}
\ee
where $\gamma^\mu$ are the Dirac matrices and the Wilson coefficients $\tilde c_d, \tilde c_s, \tilde c_v$ capture the effects of nuclear substructure obtained by integrating out its constituents. This kind of approach allows an efficient use of graphical methods in the high-energy relativistic theory (in which radiative corrections are computed) together with more efficient treatment of bound states if the nonrelativistic limit for $\Phi$ is used.  

Of more interest here are versions of these theories for which the heavy compact source is treated in a first-quantized form.  In this case both photon and electron are still described by second quantized fields, $A_\mu$ and $\Psi$, but the multi-particle states of the field $\Phi$ are integrated out, leaving only the collective coordinates --- such as the centre-of-mass position and spin -- of the single nucleus relevant to the atom of interest. Writing the worldline of the nucleus as ${\mathcal P}: x^\mu = y^\mu(s)$ the part of the effective action containing nuclear structure effects is instead written as $S=S_{QED}+ S_{Nucl}$ with:
\be
S_{Nucl}=-\int_{\mathcal P} \exd s\left\{\sqrt{-\dot y^2}M-Ze\dot y^\mu A_\mu + c_s\sqrt{-\dot y^2}\left(\overline \Psi \Psi\right)+i c_v \dot y^\mu \left(\overline \Psi \gamma_\mu \Psi\right)+\cdots \right\} \,,
\ee
where $\dot y := dy/ds$ and $c_s, c_v$ are again Wilson coefficients. In this case the path integral defining the EFT is over the fields $A_\mu, \Psi$ and the nucleus coordinate $y^\mu$ rather than $A_\mu, \Psi$ and $\Phi$. To emphasize the first-quantized nature this approach can be called the Point Particle Effective Field Theory, or PPEFT for short. 

One variant of the PPEFT formalism has been used to explore how massive objects like black holes or neutron stars interact gravitationally \cite{Goldberger:2004jt} (such as when inspiralling pairs emit gravitational waves). This variant typically works in an expansion about flat space and so profits from  the efficient graphical expansions this allows in the weak-field limit. We do not use this formulation here because position-space methods are more convenient for our purposes, as we next describe.

\subsection{Point Particle Effective Field Theory}
\label{App:PPEFT}

We here follow the position-space PPEFT approach of \cite{Burgess:2016lal, Burgess:2016ddi, Burgess:2017ekj} that is better suited to describing classical solutions in a way not tied to weak-field expansions. In this approach classical bulk equations are solved exactly and control over calculations far from any sources is maintained despite the presence of singular near-source configurations by carving out a small Gaussian pillbox of proper radius $\epsilon$ that excises each source from the surrounding bulk spacetime. The influence of the sources on the surrounding fields is then captured through appropriate boundary conditions applied at the boundaries of these pillboxes (in much the same way that Gauss' law specifies the amount of electric flux at the boundary of a pillbox in terms of the net charge contained within). 

This procedure allows the larger bulk properties to be understood without needing to know microscopic details about the sources, much as we can understand atomic energy levels in detail without first completely understanding nuclei.\footnote{The utility of the PPEFT formalism has been tested in some detail in calculations of how finite nuclear size influences precision calculations of atomic energy levels \cite{Burgess:2017mhz, Plestid:2018qbf, Zalavari:2020tez, Burgess:2020ndx}. } It also shows how to identify the boundary conditions as a function of the effective couplings in the EFT describing the `point-like' source, allowing successively more accurate boundary conditions as more and more details (like higher order multipoles) of the source become known.

More specifically, if the physical linear size, $a$, of the source is much smaller than the typical extra-dimensional length scale, $\ell$, then one chooses $\epsilon$ to lie within the regime $a \ll \epsilon \ll \ell$ for which a multipole-like expansion of bulk fields obtained as powers of $a/\epsilon$ can be matched to the boundary conditions satisfied by the bulk fields in their asymptotic near-source limit $\epsilon \ll \ell$. PPEFT methods also show how renormalizations of these effective couplings (even at the classical level along the lines seen in \cite{Goldberger:2001tn}) give renormalization-group arguments that ensure nothing physical depends on arbitrary features like the precise value of the size $\epsilon$ of the pillboxes, leaving only predictions organized by ratios of the physical quantity $a/\ell$  (for a review see \cite{EFTBook}). 

Although PPEFT arguments usually start with the EFT for the source and infer the near-source boundary conditions that follow from it, running this argument backward also allows one to learn what kinds of sources are consistent with the asymptotic near-source form of known bulk solutions (like the ones described in previous sections). This gives useful but limited information because at low energies the bulk fields only carry partial information about the nature of the source, similar to -- though a generalization of -- the information carried about electromagnetic moments in the multipole expansion of the field due to a charge distribution. 

Electromagnetic interactions provide the simplest place to start, so consider a point-like static electromagnetic source in $D$ spacetime dimensions giving rise to a bulk electromagnetic field with naive action
\be
   S_b = \int \exd^D x \; \int_W \exd s \; L_b[y(s)] \, \delta^D[x- y(s)] = \int \exd^{D}x \;\frac{L_b}{ \gamma }\, \delta^{D-1}[\bfx - \bfy(s)] \,,
\ee
where the integral is along the world-line $W$ parameterized as $x^\ssM = y^\ssM(s)$ along which $s$ is an arbitrary parameter. We here use the physicists' crutch: write the source action proportional to a delta function that localizes it at the correct position. (Part of the point of PPEFT methods is to make this crutch more precise.) The last equality uses one of the delta functions to perform the $s$ integral and defines $\gamma := \partial y^0/\partial s$. The delta function evaluates $s$ at $s = s(x^0)$. 

In the usual treatment of delta-function sources the equation of motion for this action is integrated over a small spatial Gaussian pillbox, $P_\epsilon$, of radius $\epsilon$ which the matter contribution can be performed using the spatial delta function. The integral over the bulk part of the field equation only picks up the surface term if the pillbox is small enough. For instance, if the bulk Maxwell action is $\cL = - \frac14 \sqrt{-g}\, f(\phi) F_{\ssM\ssN} F^{\ssM\ssN}$ this leads to 
\bea \label{NaiveGaugePPEFT}
  0 &=& \int_{P_\epsilon} \exd^{D-1}x \left\{ - \partial_\ssM \Bigl[ \sqrt{-g} f(\phi) F^{\ssM\ssN}  \Bigr] + \sqrt{-g} \, \nabla_\ssM \Bigl[ f(\phi) \, F^{\ssM\ssN} \Bigr] +\frac{1}{\gamma} \frac{\partial L_b}{\partial A_\ssN} \delta^{D-1}[\bfx - \bfy(s)] \right\} \nn\\
 &\simeq&  -\oint_{\partial P_\epsilon} \exd^{D-2}x \; n_\ssM \Bigl[ \sqrt{-g} \; f \,  F^{\ssM\ssN} \Bigr]  + \frac{1}{\gamma} \left( \frac{\partial L_b}{\partial A_\ssN} \right)_{\bfx=\bfy}\,,
\eea
where $n_\ssM$ is the outward-pointing unit normal vector on the pillbox surface. This returns Gauss' law if we specialize to the time direction ($N=0$) and assume the source action for a point charge: $S_b = Q \int \exd x^\mu A_\mu$.  

More generally, \pref{NaiveGaugePPEFT} also applies if $S_b$ is not simply linear in $A_\ssM$. If so then the delta-function term can depend on other bulk fields. Because these fields often diverge -- as does the Coulomb field -- at the source's position the delta-function terms of \pref{NaiveGaugePPEFT} can be ill-defined. This can be regulated by substituting the field at a distance $\epsilon$ from the source as a proxy for the on-source field, with any divergences as $\epsilon \to 0$ being renormalized into the effective couplings in the action $S_b$. In this way the small pillbox radius $\epsilon$ plays double duty: for the total divergence terms it provides a surface near (but outside) the source on which the external bulk field's flux can be measured, while for the delta-function terms it provides a regularization for the fields that strictly speaking would diverge if evaluated directly at the source position. This ability to fill both roles relies on choosing $\epsilon$ much smaller than the size of the extra dimensions but also much larger than the physical size of the source itself. 

A precise way to do all of this is to replace the delta-function formulation with a codimension-1 surface that consists of the world-tube swept out by the surface of the Gaussian pillbox (see for example \cite{Burgess:2008yx}). Rather than specifying the action at a point we can specify the action on this surface and consider the interior of the surface to be empty space. Then smoothness at the origin provides a unique boundary condition deep in the interior and the effects of the action on the codimension-one surface can be computed in a standard way using Israel junction conditions \cite{Israel:1966rt}, leading to a prediction for the boundary conditions just outside the pillbox. 

One might worry that this seems to be a very bespoke construction; why doesn't it build in too many model-dependent details? In principle the answer is that as long as a low-energy expansion is possible for these pillbox boundary conditions then it automatically captures {\it any} UV system within its domain of validity. It doesn't matter if a particular realization is used to derive what these are. In practice the relationship between the action on the surface and the initial point particle action on the particle world-line can in general be complicated \cite{Plestid:2018qbf}, but is simple in the special case of spherical symmetry because they are then just related by dimensional reduction: $S_{\rm cod 1} = \Omega_\epsilon S_b$, where $\Omega_\epsilon$ is the area of the surface of the Gaussian pillbox. 

In the end, the near-source boundary conditions that emerge for spherically symmetric sources in the case of a codimension-two object (a particle in 3 spactime dimensions or a 3-brane in 6 spacetime dimensions) is then easy to state for bulk scalar, Maxwell and gravitational fields \cite{Bayntun:2009im}. For a Maxwell field it is simply the codimension-two version of \pref{NaiveGaugePPEFT}:
\begin{equation} \label{GaugePPEFT}
   \oint_{C_\epsilon} \exd \theta \;
   \Bigl[ n_\ssN \sqrt{-g} \; f(\phi) \,  F^{\ssN\ssM} \Bigr]
  = -  \left( \frac{\delta S_b}{\delta A_\ssM} \right)_\epsilon \,,
\end{equation}
where in the applications of interest here $C_\epsilon$ is the circumference of the circle of the Gaussian pillbox in the two extra dimensions and the subscript on the right-hand side is meant to emphasize that fields are regulated by evaluating them at $r = \epsilon$ (if the source position is $r = 0$). For scalar fields residing in the bulk with kinetic energy $\cL = - \frac12 \kappa^{-2} \sqrt{-g}\, \cG_{\ssA\ssB} \partial^\ssM \phi^\ssA \, \partial_\ssM \phi^\ssB$ coupled to a codimension-two brane through a brane action $S_b[\phi]$ the analogous desired relation is 
\begin{equation} \label{ScalarPPEFT}
 \oint_{C_\epsilon} \exd \theta \;
 \left[ \frac{1}{\kappa^2} \sqrt{-g} \, \mathcal G_{\ssA \ssB} n_\ssM
 \partial^\ssM \phi^\ssB \right]
 =  - \left( \frac{\delta S_b}{\delta \phi^\ssA} \right)_\epsilon \,,
\end{equation}
where the integration is again about a small circle of proper radius $r$ encircling the source situated at $r = 0$ and the right-hand side is again evaluated at $r = \epsilon$. 

Finally, we quote the metric matching condition in the special case that the metric near a source can be written in the form $\exd s^2 = \exd r^2 + g_{ij} \, \exd x^i \, \exd x^j$ where $r$ is the proper distance away from the brane, and the surface of the gaussian pillbox is at fixed $r$. In this case this surface has extrinsic curvature $K_{ij} = \frac12 \partial_r g_{ij}$ and the metric boundary condition becomes 
\begin{equation} \label{MetricPPEFT}
   \lim_{r \to 0} \oint_{C_\epsilon} \exd \theta \;
    \left[ \frac{1}{2\kappa^2} \, \sqrt{-g} \;
   \left( K^{ij} - K g^{ij} \right) - \hbox{(flat)}\right] = -
   \; \left( \frac{\delta S_b}{\delta g_{ij}} \right)_\epsilon \,,
\end{equation}
where $K_{ij}$ is the extrinsic curvature of the fixed-$r$ surface, for which the local coordinates are those appropriate for
surfaces of constant $r$: $\{ x^i , i = 0,1,\cdots,4 \}$. The final terms denoted `flat' are the same result evaluated near the origin of a
space for which the location $r = 0$ is nonsingular. 

\begin{figure}[h!] 
\begin{center} 
\includegraphics[scale=0.5, trim = 0.5cm 0cm 1cm 0cm, clip]{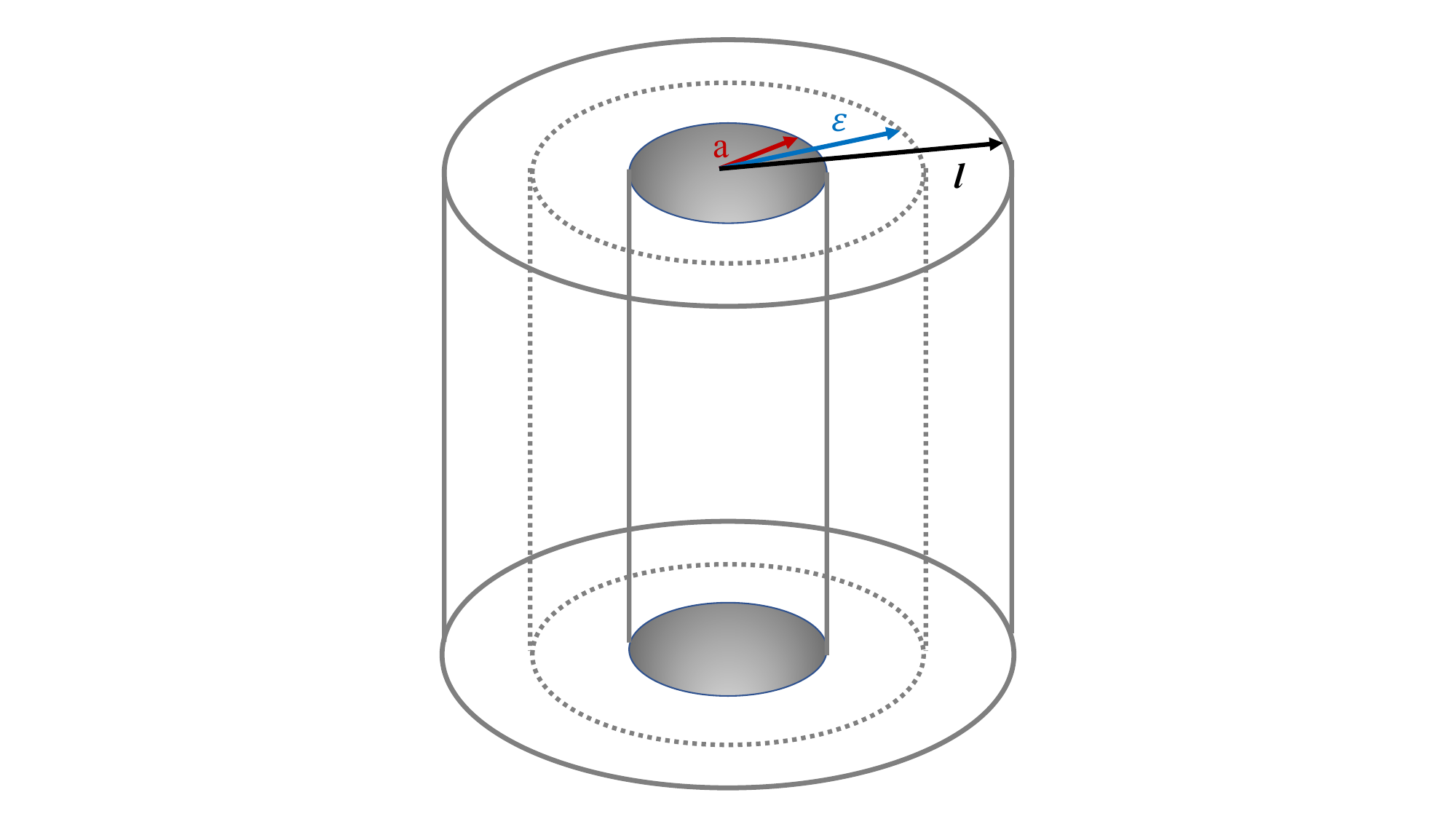}
\caption{\footnotesize Standard pillbox procedure to illustrate PPEFT. A cylindrical local source of radius $a$ is surrounded by the pillbox surface of radius $\varepsilon$. Effective field theory is valid in the bulk at a distance $\ell$ from the centre as long as $a\ll\varepsilon\ll \ell$.}
\end{center} 
\end{figure}

It should be noted that this is not that useful an equation in the special case that $i$ and $j$ lie in the $\theta$ direction that runs along the circumference of the circle $C_\epsilon$, since then it is not clear what to use for the $g_{ij}$ dependence on the right-hand side (since $S_b$ is written as a world-surface action and not as a codimension-2 action). For instance if 
\be \label{tensionSb}
  S_b = \int_W \exd^4 x \; \sqrt{-\gamma} \; T(\phi)
\ee
for $\gamma_{ab} = g_{\ssM\ssN} \partial_a x^\ssM \partial_b x^\ssN$ the induced metric on a 3-brane in 6D and $T$ the source tension, then its dependence on {\it e.g.}~$g_{\theta\theta}$ is not specified. This is not in practice a problem, however, because the Einstein equations in the $r$-$r$ direction impose a constraint on the initial data that can be used when integrating in the $r$ direction and it is this constraint that in practice fixes quantities like $\delta S_b/\delta g_{\theta\theta}$ (see below for more details in concrete examples).

Eqs.~\pref{GaugePPEFT}, \pref{ScalarPPEFT} and \pref{MetricPPEFT} can each be used in one of two complementary ways:
\begin{itemize} 
\item The most straigtforward way is to regard $\epsilon$ and the couplings -- like $T[\phi(\epsilon)]$ in \pref{tensionSb} -- to be specified, in which case these equations furnish boundary conditions that help determine the integration constants for the solutions to the field equations in the bulk, outside all of the pillboxes. Physical predictions ({\it e.g.}~bulk energy levels or scattering rates for bulk fields from the sources) depend on these integration constants and it is through their dependence on the boundary conditions that the bulk fields learn about the presence of the sources.

\item The other way to read eqs.~\pref{GaugePPEFT}, \pref{ScalarPPEFT} and \pref{MetricPPEFT} is as Callan-Symansik equations that tell us how the effective couplings, $c_i$, in $S_b$ must depend on $\epsilon$ in order to ensure that the integration constants (and so also physical observables) remain unchanged if we vary $\epsilon$ \cite{Burgess:2016lal, Burgess:2016ddi, Burgess:2017ekj}. Observables should remain unchanged as we do so because the size of the Gaussian pillboxes are arbitrary after all, and so their precise positions should not matter. The ability to see explicitly why this happens is indeed a nice feature of the PPEFT formalism. Physical predictions end up depending only on the RG invariants that characterize the entire RG trajectory $(\epsilon, c_i(\epsilon))$, rather than on the effective couplings $c_i(\epsilon)$ and $\epsilon$ separately.  
\end{itemize}

\subsection{Source properties for 6D de Sitter solutions}

In this section we follow \cite{Bayntun:2009im} and use eqs.~\pref{GaugePPEFT} through \pref{MetricPPEFT} to match the local near-brane asymptotic behaviour found in the solutions of \S\ref{Sec:OldSchool} to properties of the source action (see also \cite{Burgess:2007vi, Burgess:2008yx, Peloso:2006cq, deRham:2007mcp, Papantonopoulos:2006dv, Yamauchi:2007wm, Kaloper:2007ap, Appleby:2007ct, Arroja:2007ss}). We assume cylindrically symmetric bulk solutions and keep only the leading no-derivative terms of the source action, which we assume has the form
\be \label{BraneAction}
  S_b = - \int \exd^4 x \sqrt{-\gamma} \;  L_b(\varphi) = - \int \exd^4 x \sqrt{- g} \; W^4_b L_b(\varphi) = - \int \exd^4 x \sqrt{- g} \; T_b (\varphi)\,,
\ee
where  $\gamma_{\mu\nu} = \hat g_{\ssM\ssN} \, \partial_\mu x^\ssM \partial_\nu x^\ssN$ is the induced metric on the source, which for an unbent static source is simply the 4D metric components, $\hat g_{\mu\nu}$, from \pref{6DMetricAnsatz}. 

The last equality in \pref{BraneAction} simply defines $T_b(\varphi) = W^4_b\, L_b(\varphi)$ where $W_b = W(x_b)$ is the warp factor evaluated at the brane position. In general functions like $W$, $\varphi$ or $a$ might vanish or diverge at the brane position (if this is idealized as being pointlike) and the PPEFT approach regularizes this by replacing them by their values at the surface $r = \epsilon$ of the small Gaussian pillbox.

The resulting matching conditions are summarized in eqs.~\pref{GaugePPEFT}, \pref{ScalarPPEFT} and \pref{MetricPPEFT}, which specialized to the action \pref{salamsezgin} implies the following near-source matching relation for the scalar $\varphi$:
\be \label{scalarmatchingNS}
 \frac{2\pi}{\kappa^2} \, \Bigl[  a W^4 \, \partial_r \, \varphi \Bigr]_{r=\epsilon}  = \frac{\partial}{\partial\varphi} \Bigl[\; W_b^4 \, L_b \Bigr]
 \quad \hbox{and so} \quad \lim_{\eta \to \pm \infty} \Bigl[ \mp
  \partial_\eta \, \varphi \Bigr]  = q_\pm  = \frac{\kappa^2}{2\pi} \, \left(
  \frac{\partial \, T_\pm}{\partial\varphi} \right) \,,
\ee
where $r$ denotes proper distance away from the source and so satisfies
\be \label{GGPpropdist}
    \exd r := \mp a \, W^4 \, \exd \eta \,,
\ee 
with the sign corresponding to whether or not $\exd \eta$ points towards or away from the source in the two asymptotic regions. 

The corresponding expressions for $a$ and $W$ come from the matching conditions for different components of the metric, and for the geometry of \pref{6DMetricAnsatz} the $(\mu\nu)$ components of the metric matching
conditions are
\begin{equation} \label{munumatchingW}
 - \frac{2\pi}{\kappa^2} \left\{ W^4 \left[ a \left( 3  \frac{ \partial_r W}{W}  + \frac{\partial_r a}{a} \right)  - 1 \right]
 \right\}_{r=\epsilon} = W_b^4 L_b(\varphi) \,,
\end{equation}
and so
\begin{equation} \label{munumatchingW}
  \lim_{\eta \to \pm \infty} \left\{ \mp \left[ 3 \left( \frac{ \partial_\eta W}{W} \right)
 + \left( \frac{ \partial_\eta a}{a} \right) \right] - W^4
 \right\} = 3 \omega_\pm + \alpha_\pm - W_\pm^4  = - \frac{\kappa^2 T_\pm}{2\pi}  \,.
\end{equation}
This is a familiar expression in the special case where $\omega_b = 0$, in which case $W$ asymptotes to a constant $W \to W_b$ in the near-source limit. Using the asymptotic expression $a \simeq a_b \, e^{-  \alpha_b |\eta|}$ implies the near-source extra-dimensional geometry of \pref{6DMetricAnsatz} becomes
\be
 a_b^2 e^{- 2\alpha_b|\eta|}(W_b^8 \exd \eta^2 + \exd \theta^2 ) \simeq \exd r^2 + \left( \frac{r \alpha_b }{W_b^4} \right)^2 \exd \theta^2 \,,
\ee
revealing a near-source conical singularity with defect angle that \pref{munumatchingW} states has size 
\be
 \delta_b := 2\pi \left( 1 - \frac{\alpha_b}{W_b^4} \right) = \frac{\kappa^2 T_b}{W_b^4} = \kappa^2 L_b \,,
\ee
in agreement with standard formulae.

The matching condition that comes from the $( \theta\theta)$ component of the metric gives 
\be \label{thetathetamatchingW}
 \frac{2\pi}{\kappa^2} \Bigl[ a W^4 \, \partial_r \, W  \Bigr]_{r = \epsilon} = W^4_b \tilde U_b(\varphi) = U_b(\varphi)
 \quad \hbox{and so} \quad \lim_{\eta \to \pm \infty} \left[ \mp  \left( \frac{\partial_\eta \, W}{W}
 \right) \right] = \omega_\pm = \frac{\kappa^2 U_\pm}{2\pi} \,.
\ee
where $U_b$ and $\tilde U_b$ are defined by $U_b = - \frac12 \partial T_b/\partial g_{\theta\theta}$ and $\tilde U_b = - \frac12 \partial L_b/\partial g_{\theta\theta}$. At first sight this seems not so useful because the dependence of the source action on $g_{\theta\theta}$ is usually not specified in expressions like \pref{BraneAction} for the action of a pointlike source in the extra dimensions. But a dependence on $g_{\theta\theta}$ is implicit in any microscopic description of a source, and the good news is that this dependence is imprinted in the bulk field equations through the $(\eta\eta)$ Einstein equation that gives the constraint \pref{einsteinconstraint}, evaluated near the source. 

To see what this implies explicitly, drop terms in \pref{einsteinconstraint} that are subdominant in powers of $r$ in the near-source limit. This yields ({\it c.f.}~eq.~\pref{qconstraint})
\bea \label{einsteinconstraintU}
    0 &\simeq&  \left[ - \frac{8\,a'W'}{aW} - \frac{12(W')^2}{W^2} +  (\varphi')^2  \right]_b =  - 8\alpha_b \omega_b - 12\omega_b^2 + q_b^2    \nn\\
    &=& - 8\, \cU_b \Bigl(  W^4_b - \cT_b - 3 \cU_b \Bigr) - 12 \,\cU_b^2 + (\cT_b{\,'})^2 \nn \\
    &=& W_b^8 \Bigl[ - 8\, \tilde\cU_b \Bigl( 1 - \cL_b - 3 \tilde \cU_b \Bigr) - 12 \, \tilde \cU_b^2 + (\cL_b{\,'})^2 \Bigr]\,,
\eea
where the prime on $\cT_b{\,'}$ denotes differentiation with respect to $\varphi$. The second line uses the matching conditions \pref{scalarmatchingNS}, \pref{munumatchingW} and \pref{thetathetamatchingW} and defines the dimensionless quantities $\cT_b := \kappa^2 T_b/(2\pi) = W_b^4 \cL_b$ -- with $\cL_b := \kappa^2 L_b/(2\pi)$ -- and $\cU_b := \kappa^2 U_b/(2\pi) = W^4_b \tilde \cU_b$ -- where $\tilde \cU_b := \kappa^2 \tilde U_b/(2\pi)$. The point is that eq.~\pref{einsteinconstraintU} can be solved to get $\cU_b$ as a function of $\cL_b$ and its derivative $\cL_b'$, 
\be \label{Ubsoln}
 \cU_b = W_b^4 \tilde \cU_b = \frac{W_b^4}3 \left[ (1 - \cL_b) - \sqrt{(1 - \cL_b)^2
 - \frac34 \, (\cL_b{\,'})^2} \right]  \,,
\ee
both of which {\it can} be read off from \pref{BraneAction}. 

The root in \pref{Ubsoln} is chosen so that the result  for $\cU_b$ vanishes when $\cL_b{\,'} = 0$, since in this case the source couplings do not break the shift symmetry \pref{SSRescaling} and so cannot lift the degeneracy in \pref{SSdegeneracy}. Indeed ref.~\cite{Bayntun:2009im} explicitly computes the scalar potential for this would-be flat direction in the effective theory obtained by compactifying to 4D, with the result
\be
 V_{\rm\,eff}(\varphi_0) = - \sum_b \left( U_b + \frac{T_b'}{2} \right) = - \sum_b W_b^4 \left(\tilde U_b + \frac{L_b'}{2} \right) \,,
\ee
ensuring that the Friedmann equation computed in the 4D EFT agrees with \pref{IntegratedCBA-GGP} once the matching conditions  \pref{scalarmatchingNS}, \pref{munumatchingW} and \pref{thetathetamatchingW} are used. Using \pref{Ubsoln} shows that this potential necessarily vanishes when $T_b{\,'}$ vanishes for both sources, corresponding to the flat direction \pref{SSdegeneracy}. 

\pagebreak

\section{Two stringy routes to 6D supergravity}
\label{Sec:10DUplift}

\begin{quote}
\rightline{{\it When you come to a fork in the road, take it.} } 
\end{quote}

\noindent
\S\ref{Sec:OldSchool} describes in detail exact 4D de Sitter solutions to 6D $(1,0)$ gauged supergravity. It remains to lift these to solutions of field equations with well-established M-theory pedigrees, which requires knowing how 6D $(1,0)$ supergravity can be obtained from higher dimensions. At present there are two known ways to do so:  
\begin{enumerate}
\item {\it Consistent truncation of 10D heterotic/Type I supergravity:} 6D chiral supergravity can be obtained by truncating the ten dimensional heterotic/type I supergravity on ${\mathcal H}^{(2,2)} \times S^1$ where ${\mathcal H}^{(2,2)}$ is a three-dimensional hyperbolic manifold and $S^1$ is a circle \cite{Cvetic:2003xr}. This is a consistent truncation inasmuch as it is consistent with the classical equations of motion.
\item {\it Flux F-theory compactification on elliptically fibered Calabi-Yau: } 6D chiral supergravity can be obtained from 10D type IIB supergravity via an $F$-theory reduction on an elliptically fibered Calabi-Yau manifold manifod \cite{Grimm:2013fua}.
\end{enumerate}
Although our main focus is on the second of these approaches, for completeness' sake this section briefly reviews both of them.

\subsection{Type I/Heterotic uplifts}

In this section we summarise the claims of \cite{Cvetic:2003xr}, and return to the F-theory approach in \S\ref{ssec:FtheoryUplift}.

\subsubsection*{String theory on Hyperbolic space ${\mathcal H}^{2,2}\times S^1$}

In this approach the 6D theory is obtained by a series of dimensional reductions and consistent truncations. At the end it amounts to  a dimensional reduction of the effective action of type I/heterotic strings on ${\mathcal H}^{(2,2)}\times S^1$ where ${\mathcal H}^{(2,2)}$ is  the 3D hyperbolic space determined by the real coordinates $y_i$, $i=1,\cdots, 4$ subject to the constraint $y_1^2+y_2^2-y_3^2-y_4^2=1$ and $S^1$ a circle of radius $R$.

We start with the relevant  heterotic/type I bosonic part of the Lagrangian (where hats denote 10D quantities):
\be   \label{tend}
    \cL_{10} = - \sqrt{-\hat g} \left[ \frac{1}{2 \kappa_{10}^2} \,
    \hat g^{\ssM\ssN} \Bigl( \hat R_{\ssM\ssN} + \partial_\ssM \hat\varphi \,
    \partial_\ssN \hat\varphi \Bigr) + \frac{1}{4}  e^{-  \hat\varphi} \hat F_{\ssM\ssN} \hat F^{\ssM\ssN}+ \frac{1}{12}  e^{-  2\varphi} \hat H_{\ssM\ssN\ssP} \hat H^{\ssM\ssN\ssP}
     \right] \,,
\ee
The idea is to compactify on the 4D space ${\mathcal H}^{(2,2)}\times S^1$. To this end we (in this section) adopt Planck units ($\kappa_{10} = 1$) and denote the coordinates of ${\mathcal H}^{(2,2)}$ by $\rho, \alpha, \beta $, where 
\begin{equation}
y_1+i y_2 =\cosh\rho e^{i\alpha}; \qquad y_3+i y_4 =\sinh\rho e^{i\beta}
\end{equation}
with ranges $0\leq \rho \leq \infty$, $0\leq \alpha,\beta < 2\pi$. Finally, use $z$ as the $S^1$ coordinate. The solutions found in  \cite{Cvetic:2003xr} correspond to the metric:
\begin{equation}
\exd\hat s^2_{10}=\left(\cosh 2\rho \right)^{1/4} \left[ e^{-\varphi/4}\exd s_6^2+e^{\varphi/4}\exd z^2+\frac{e^{\varphi/4}}{2\bar g^2}\left( \exd\rho^2 + \frac{\cosh^2\rho}{\cosh 2\rho}\left(D\alpha\right)^2 + \frac{\sinh^2\rho}{\cosh 2\rho}\left(D\beta \right)^2 \right) \right]
\end{equation}
where $D\alpha := \exd\alpha-\bar g A_{(1)}$ and $D\beta := \exd\beta +\bar g A_{(1)}$.

The antisymmetric field similarly takes the form:
\begin{equation}
\hat H_{(3)}=H_{(3)}+\frac{\sinh2\rho}{2\bar g (\cosh 2\rho)^2}\exd\rho \wedge D \alpha\wedge D \beta+\frac{1}{2\bar g \cosh 2\rho}F_{(2)}\wedge\left(\cosh^2\rho\, D\alpha-\sinh^2\rho\, D \beta\right)
\end{equation}
where $F_{2}= \exd A_{(1)}$ and the field strength $H_{(3)}$ satisfying the Bianchi identity $\exd H_{3}=\frac12 F_{(2)}\wedge F_{(2)}$.

Finally the 10D dilaton $\hat\phi$ (which from the 
{Type} I truncation of type IIA strings determines the string coupling $g_s=\langle e^{\hat\phi} \rangle $ ) relates to the 6D dilaton $\varphi$  as:
\begin{equation}
e^{\hat \phi}=\left(\cosh 2\rho \right)^{-1/2}\, e^{\varphi}
\end{equation}
so weak string coupling corresponds to large $\rho$ and/or large negative $\varphi$. For the Type I truncation of IIB strings, the same ansatz holds with the exchange of the three  form $F_{(3)}$ from NS-NS to RR  and S-duality would give the opposite sign for $\hat\phi$ and then a different interpretation of weak/strong coupling  \cite{Cvetic:2003xr}.
  
Ref.~\cite{Cvetic:2003xr} claim that the above ansatz for the metric, anti-symmetric tensor and dilaton plugged into the 10D supergravity equations gives rise to the 6D equations for  $g_{\ssM\ssN}, B_{(2)}, A_{(1)}, \varphi $ derived from the Salam-Sezgin Lagrangian (\ref{salamsezgin}). Therefore {\it any} solution of the 6D Salam-Sezgin theory can be lifted to 10D by the inverse of this procedure \cite{Cvetic:2003xr}. In particular the well known $R^{1,3}\times S^2$ solution found by Salam and Sezgin has an uplift to 10D which in the large $\rho$ limit corresponds to the linear dilaton string background. In this construction the value of the 6D Newton constant $\kappa$ is not derived in the usual way as a convergent integral of the 10D Newton constant because of the noncompact nature of the extra dimensions. In principle one can apply the same reasoning also to the Minkowski, AdS and dS solutions found in \cite{Tolley:2005nu} and thereby promote those solutions to the full 10D theory. 

\subsection{F-theory uplifts}
\label{ssec:FtheoryUplift}


Let us now discuss the second derivation of the 6D $(1,0)$ supergravity from string theory. This follows the by-now standard treatment of defining F-theory constructions in terms of M-theory compactifications\footnote{See for instance the detailed discussion in \cite{Denef:2008wq}.} to one less dimension. We start  with 11D supergravity compactified on a six-dimensional elliptically fibered Calabi-Yau manifold $Y_6$ which is locally a product of a 4D basis space $B_4$ and a two-torus $T^2$. $T^2$ is itself a product of two circles $S^1_A$ and $S^1_B$.  The 11D manifold can be written as:
\be
{\mathcal M}_{11}= {\mathbb R}^{1,4}\times Y_6; \qquad Y_6\simeq B_4\times T^2; \qquad T^2=S^1_A \times S^1_B
\ee
Compactifying on the circle $S^1_A$ gives rise to type IIA string theory. T-dualising on the circle $S^1_B$ ($S^1_B\rightarrow {\tilde S}^1_B$ with the radius of ${\tilde S}^1_B$ being large) gives rise to the IIB string theory compactification for which the 10D manifold is locally:
\be
{\mathcal M}_{10}= {\mathbb R}^{1,4}\times {\tilde S}^1_B\times B_4 \rightarrow  {\mathbb R}^{1,5}\times B_4 
\ee
In the last step we take the large radius limit of the circle ${\tilde S}^1_B$ that promotes the 5D Poincar\'e symmetry to the full 6D Poincar\'e symmetry.

The procedure to derive the 6D effective action is as follows:
\begin{itemize}
\item Dimensionally reduce the 11D theory to 5D on a Calabi-Yau manifold $Y_6$ to obtain a 5D ${\mathcal N}=2 $ supergravity theory.
\item Dimensionally reduce the most general 6D $(1,0)$ supergravity theory to obtain a subclass of the  5D ${\mathcal N}=2 $ supergravity theory.
\item Compare the two results above in order to identify the particular class of 6D theories that corresponds to this compactification.
\end{itemize}
Following this procedure Grimm and collaborators were able to identify the 6D supergravity theory\footnote{Contrary to the hyperbolic compactification above, we cannot claim that the F-theory realization corresponds to a consistent truncation. Actually,  there is not yet a mathematically rigorous formulation of the theory \cite{Denef:2008wq}. However, given the large amount of work illustrating concrete consistent models and the strong mathematical basis of the theory, there is little doubt of its validity. Furthermore the constructions following the M-theory prescription have reproduced a series of consistent results. The techniques followed in \cite{Grimm:2013fua} in which solutions of the 6D theory can reproduce direct compactifications to 4D further add to the believe of the consistency of these constructions.} \cite{Bonetti:2011mw,Grimm:2013fua}.

Starting with  11D supergravity with its multiplet composed of the graviton $g_{\ssM\ssN}$, gravitino $\psi_\ssM$ and antisymmetric tensor $C_{\ssM\ssN\ssP}$ with field strength $G_{(4)} =\exd C_{(3)}$ or  $G_{\ssM\ssN\ssP\ssQ}=\partial_{[\ssM}C_{\ssN\ssP\ssQ]}$. The bosonic action takes the form:
\be
{S_{11}}= -\frac{1}{2\kappa_{11}^2}\int \exd^{11}x \left[ \sqrt{-g}\left (   R+\frac{1}{48}G_{\ssM\ssN\ssP\ssQ}G^{\ssM\ssN\ssP\ssQ}\right)+\frac{2}{3\cdot 4!^3}\epsilon^{\ssM_1\cdots \ssM_{11}}C_{\ssM_1\cdots \ssM_3} G_{\ssM_4\cdots \ssM_7}G_{\ssM_8\cdots \ssM_{11}}\right] \,.
\ee
Upon compactification to 5D, the metric can be split into the non-compact 5D and the compact 6D corresponding to the Calabi-Yau space $Y_6$:
\be
 \exd s^2= \exd s_5^2+ \exd s_6^2 \,.
\ee
Turning on fluxes for $G_{\ssM\ssN\ssP\ssQ}$ in four of the internal dimensions  corresponding to the non-trivial four-cycles of $Y_6$ through the Freund-Rubin ansatz:
\be
\int_{\gamma_\Lambda} \langle G_{(4)} \rangle=n_\Lambda;\qquad \langle G_{(4)}\rangle=n^\Lambda\omega_\Lambda
\ee
where $\omega_\Lambda$,  $\Lambda=1,\cdots , h_{1,1} $ are a basis of four-forms associated to the four-cycles $\gamma_\Lambda$ in the Calabi-Yau manifold $Y_6$. With $h_{1,1}$ the corresponding Hodge numbers and $n_\Lambda$ are  integers. As usual, plugging these fluxes back into the action gives rise a positive scalar potential for the corresponding moduli.

With this background for the bosonic 11D fields the 5D massless degrees of freedom are identified in terms of the fluctuations of the metric $\delta g_{\ssM\ssN}$ and antisymmetric tensor $\delta C_{\ssM\ssN\ssP}$ as follows:
\begin{itemize}
\item {\it Metric}: The 5D metric $g_{\mu\nu}$, $\mu,\nu= 1, \cdots, 5$. The  K\"ahler moduli $v^\Lambda$, $\Lambda=1,\cdots , h_{1,1} $, corresponding to the (real) sizes of the four-cycles and the  (complex) complex structure moduli $z_\kappa$, $\kappa=1,\cdots, h_{1,2}$

\item{\it Antisymmetric tensor}: The 5D antisymmetric tensor $C_{\mu\nu\rho}$ which is dual to a 5D scalar $\Phi$.  The fluctuations of the internal components  $C_{mnp}$ corresponding to scalars $\xi^K, \tilde\xi_K$ with $K=0,1, \cdots, h_{1,2}$. There also the fluctuations of $C_{mn\mu}$ which correspond to $h_{1,1}$ 5D vectors $A_\mu^\Lambda$.
\end{itemize}

All of these massless fields can be accommodated into 5D multiplets with bosonic components  as follows:
\begin{itemize}
\item{\it Gravity multiplet}: The 5D graviton $g_{\mu\nu}$ and the graviphoton corresponding to one of the vector fields $A_\mu^\Lambda$.
\item{\it Vector multiplets}: The remaining $h_{1,1}-1$ vectors $A_\mu^\Lambda$ together with the  $h_{1,1}-1$ constrained fields $L^\Lambda$ determined by the K\"ahler moduli $v^\Lambda$.
 \be
L^\Lambda=\frac{v^\Lambda}{{\mathcal V}^{1/3}},\qquad {\mathcal N}\equiv \lambda_{\Lambda\Omega\Gamma}L^\Lambda L^\Omega L^\Gamma=1
\ee
where $\cV$ is the overall volume (in string units) of the Calabi-Yau $Y_6$, ${\mathcal V}=\lambda_{\Lambda\Omega\Gamma} v^{\Lambda}v^\Omega v^\Gamma$ with $\lambda_{\Lambda\Omega\Gamma}$ the triple intersection numbers:
\be
\lambda_{\Lambda\Omega\Gamma}=\int_{Y_6}\tilde\omega_\Lambda\wedge \tilde\omega_\Omega\wedge \tilde\omega_\Gamma.
\ee
with $\tilde \omega_\Lambda$ the dual of $\omega_\Lambda$ in 6D, etc.
\item {\it Hypermultiplets}: $q^u$, $u=1,\cdots, 4(h_{1,2}+1)$ corresponding to the fields $\left({\mathcal V}, \Phi, z^\kappa,\bar z^{\bar\kappa},\xi^K, \tilde \xi_K\right)$.   The volume $\cV$, together with $\Phi$ and $\xi^0, \tilde\xi^0$ compose the universal hypermultiplet. In F-theory, one combination of the $z^\kappa$ corresponds to the string dilaton $\tau$.
\end{itemize}

With this information the effective action for these fields has been explicitly computed. The relevant effect from the fluxes correspond to the kinetic term for the field $\Phi$ which is written in terms of a covariant derivative $D\Phi$ and an effective positive flux potential $V_{flux}$:
\begin{equation} \label{VfluxForm}
D_\mu\Phi=\partial_\mu \Phi + 2 n_\Lambda A_\mu^\Lambda;\qquad V_{flux}=\frac{1}{8{\mathcal V}^2}G^{\Lambda\Sigma}n_\Lambda n_\Sigma
\end{equation}
where $G^{\Lambda\Sigma}$ is the inverse of the metric $G_{\Lambda\Sigma}(L)$ determining the kinetic terms for the fields $L^\Lambda$: 
\be
G_{\Lambda\Sigma}=\frac{1}{2{\mathcal V}^{1/3}}\left(\int_{Y_6} \tilde w_\Lambda \wedge  w_\Sigma\right).
\ee

\subsubsection{6D uplift and equations of motion}

The 5D theory can be uplifted (or oxidized) to find the corresponding 6D action. The result is a particular case of the most general 6D theory. 

For these purposes it is convenient to split the components of the 5D vector multiplets into three classes depending on the nature of the corresponding $h_{1,1}(Y_6)$ moduli $v^\Lambda$ by splitting the indices $\Lambda$ as $\Lambda=\{0,\alpha,i\}$ with $v^0$ corresponding to the four-cycle of $Y_6$ that is the full $B_4$; $v^\alpha$ corresponding to the 4-cycles of $Y_6$ dual to the 2-cycles of $B_4$ so $\alpha=1, \cdots, h_{1,1}(B_4)$ and $v_i$ corresponding to the 4-cycles that combine the fibre and the base. When the $v^i$ vanish it corresponds to co-dimension two singularities that are identified with D7 branes in the weak coupling limit. The gauge group is usually non-abelian but we will restrict to the Coulomb branch and then $i=1, \cdots, {\rm rank}\, G$.

The 6D to 5D dictionary corresponds to:
\be
n_H=h^{2,1}(Y_6)+1; \qquad n_T+1= h^{1,1}(B_4);\qquad {\rm rank}\,G=h^{1,1}(Y_6)-h^{1,1}(B_4)-1
\ee
Since we are restricting to $n_\ssT=1$ this implies we are considering backgrounds with $ h^{1,1}(B_4)=2$ and so $h^{1,1}(Y_6)\geq 3$.

In the weak coupling limit, the $G_{mnpq}$ fluxes correspond to magnetic fluxes from D7 branes $\int_{\gamma_\Lambda} F_{(2)}=n^\Lambda$ with $\gamma_\Lambda$ a two-cycle wrapped by the D7 brane and the scalar potential is obtained from the kinetic term in the D7 DBI action: 
\be 
\langle F_{(2)}\rangle=n^\Lambda \tilde\omega_\Lambda; \qquad \int \exd^8 x \sqrt{-g}F^{MN}F_{MN}\implies V_{\rm D7}\propto G^{\Lambda\Sigma}n_\Lambda n_\Sigma
\ee
and the gauging of the field $\Phi$ comes from the Chern-Simons coupling $\int  C_{(4)}\wedge F_{(2)}\wedge F_{(2)}$ with $C_{(4)}$ the RR four-form of type IIB and $F_{(2)}$ the corresponding gauge fields, which in tensor notation is $\int \exd^8 x \, C_{\mu\nu\rho\sigma}F_{\alpha\beta}F_{mn}\epsilon^{\mu\nu\cdots mn}$. Upon dualization of $C_{\mu\nu\rho\sigma}$ to $\Phi$ this gives rise to a term $n A_\mu \partial^\mu \Phi$ in the Lagrangian that gives rise to the gauging of $\Phi$. In this case the scalars $z^\kappa$ correspond to D7 brane fluctuations and the $\xi_\ssI$ to Wilson line $A_m$ fluctuations.

We next write down the relevant equations to be solved. These are derived in general in ~\cite{Grimm:2013fua}. We concentrate on the case $n_\ssT=1$.
Since we only seek maximally symmetric solutions in 4D, we set $H^{\ssM\ssN\ssP} = 0$ from the start. Of the other bosonic fields, the metric $g_{\ssM\ssN}$,  hypermultiplets $q^\ssU$ and scalars from tensor multiplets $j^\alpha$, we consider non-trivial profiles only for:
\be
{\rm Metric}: g_{\ssM\ssN};\qquad{\rm Abelian\,  vector}: A_\ssM;\qquad {\rm  Hypermultiplets}\,\, q^\ssU: \left(\cV, \Phi\right), \tau; \qquad {\rm Tensor\, multiplets}: \varphi
\ee
with all other fields taken to be constant and satisfying their field equations trivially. Here $\tau$ is the complex dilaton field that in F-theory corresponds to a complex structure modulus.

The 6D action now is more complicated than the original Salam-Sezgin action because we cannot ignore the hypermultiplets $q^\ssU$. Then equation \eqref{salamsezgin} is generalized to:
\bea   \label{salamsezgin2}
   -\frac{ \cL_6}{\sqrt{-g} } &=&  \frac{1}{2 \kappa^2} \,
    g^{\ssM\ssN} \Bigl( R_{\ssM\ssN} + \partial_\ssM \varphi \,
    \partial_\ssN \varphi +h_{\ssU\ssV}D_\ssM q^\ssU D_\ssN q^\ssV\Bigr) \nn \\ 
    & &  + \frac{1}{4}  e^{-  \varphi} F_{\ssM\ssN} F^{\ssM\ssN}+ \frac{1}{12}  e^{-  2\varphi} H_{\ssM\ssN\ssP} H^{\ssM\ssN\ssP}
    + \frac{2\mfg^2 U(q)}{\kappa^4} \, e^{\varphi}  \,,
\eea
where the scale $\mfg$ comes from fields with nonzero fluxes in the dimensions beyond the 6D on which the fields in this lagrangian depend. The  target-space metric for the hypermultiplet fields is diagonal and takes the form \cite{Grimm:2013fua}:
\be
   h({\cV}) := h_{\cV \cV}=4h_{\Phi\Phi} =\frac{1}{2\cV^2}; \qquad h_{\tau \tau^*}=\frac{1}{2\tau_2^2}
\ee 
where $\tau=\tau_1+ \text{i}\tau_2$. The only non-trivial covariant derivative is for $\Phi$: $D_\ssM\Phi = \partial_\ssM \Phi + k A_\ssM$, where $k$ must be nonzero if the gauge field $A_\ssM$ is the same one that is responsible for the flux that generates the scalar potential in \pref{salamsezgin2}, but can vanish if the 6D gauge potential is distinct from the one whose flux is responsible for the potential.

The scalar potential descending from \pref{VfluxForm} is
\be
V(\varphi, \chi) 
=\frac{2\mfg^2 }{\kappa^4\cV^2} \, e^{\varphi}=\frac{2\mfg^2 }{\kappa^4} \, e^{\varphi-2\chi} \quad\hbox{where} \quad \cV := e^\chi \,,
\ee
and we introduce the canonically normalized field $\chi = \log \cV$ and use $U(q) \propto \cV^{-2} = e^{-2\chi}$. Notice that $V$ depends on both the tensor-multiplet scalar $\varphi$ and the hypermultiplet scalar $\cV=e^\chi$. This flux induced potential  is a particular case of the general 6D scalar potential with $U\propto 1/\cV^2$. The volume dependence is determined by the fluxes and the Weyl rescaling of the metric to get the standard Einstein frame.
Unlike the Salam-Sezgin case we have a runaway potential for $\chi$ in addition to $\varphi$.

Since we are interested in maximally symmetric solutions in 4D we restrict to backgrounds for which $H_{\ssM\ssN\ssP}=0$. The field  equations then simplify to the trace-reversed Einstein equation
\be
\label{eq:RMNeq}
R_{\ssM\ssN} = 4 e^{-\varphi} \kappa^2 F_{\ssM\ssP} F^{\ssP}_\ssN + \partial_\ssM\varphi \partial_\ssN \varphi + \frac{1}{2} \partial_\ssM \chi \partial_\ssN \chi + \frac12 e^{-2\chi}D\ssM\Phi D\ssN\Phi+\frac{1}{2 \tau_2^2} \partial_{(\ssM} \tau \partial_{\ssN )} \tau^*+  \frac{1}{2} g_{\ssM\ssN} \Box\varphi \,,
\ee
and the Maxwell equation 
\be \label{eq:Maxwell6DF}
\nabla_\ssR\left(e^{-\varphi}F^{\ssL\ssR}\right) =\frac{k}{32}{e^{-2\chi}} D^\ssL\Phi\,,
\ee
and the scalar equations
\begin{eqnarray} \label{eq:Scalars6DF}
 &&\Box \chi = -4 e^{\varphi - 2 \chi} \tilde{V} - e^{-2\chi} D_\ssM\Phi D^\ssM\Phi \,, \qquad
\Box \Phi  ={2\partial_\ssM\chi}D^\ssM\Phi \,,\\
 &&\; \Box\varphi = \tilde{V} e^{\varphi - 2 \chi} - \kappa^2 e^{-\varphi} F^{\ssM\ssN}F_{\ssM\ssN} \quad \hbox{and} \quad
\Box\tau  = - \frac{i}{\tau_2}\partial^\ssM\tau \partial_\ssM\tau \,, \nn
\end{eqnarray}
with $\tilde V \propto \mfg^2/\kappa^4$ the constant determined from higher-dimensional fluxes appearing in the 6D scalar potential $V=\tilde V e^{\varphi-2\chi}$. Notice that $\tilde V$ is the only parameter through which $\kappa$ and $\mfg$ enter into the solutions and so sets their overall scale.  In particular $\tilde V$ can be much smaller than the Planck size if $\mfg^2/\kappa \ll 1$.

We see that these equations allow for solutions for which the dilaton field $\tau$ is constant but in  F-theory, and similar to the stringy cosmic string solutions \cite{Greene:1989ya}, there can also be solutions with  non-trivial  profile for $\tau$ that we can also consider. Since  the scalar potential is runaway in the $\varphi$ and $\chi$ directions  there are no solutions for which both of these fields are constant in 6D: as before there are no maximally symmetric solutions in 6D.  

Ref.~\cite{Grimm:2013fua} finds explicit solutions to these equations assuming constant $\varphi$, which is consistent with the equations of motion if the fluxes of the gauge field are adjusted to cancel the contribution from the scalar potential on the right-hand side of the $\varphi$ equation. This solution is particularly interesting since they prove that it preserves ${\mathcal N}=1$ supersymmetry in 4D and reproduces the features of standard 4D compactifications of F-theory, but in a two-stage approach that passes through six dimensions.  Checks of the remaining EFT in the two stage (F theory to 6D to 4D) and direct (F-theory to 4D) approaches reproduce the same EFT giving robustness to the procedure.

We next consider maximally symmetric solutions in 4D but without supersymmetry for which both dS and AdS solutions exist. We do so by considering non-trivial profiles for both the volume field $\chi$ and the gauge coupling field $\varphi$, thereby generalizing both the generalized  Salam-Sezgin solutions in which only $\varphi$ was considered and the solutions of \cite{Grimm:2013fua} in which only $\chi$ was considered.

\section{4D de Sitter from F-theory derived 6D supergravity}
\label{Sec:NewWave}

\begin{quote}
\rightline{{\it It's like d\'ej\`a vu all over again.}}
\end{quote}

\noindent
We now consider maximally symmetric solutions in 4D of which there are several interesting cases:
\begin{enumerate}
\item Constant $\tau$  and $k=0$ which means the 6D flux is from a $U(1)$ field that is orthogonal to the combination that gives the Stuckelberg contribution to $D\Phi$. In this case $D_\ssM\Phi=\partial_\ssM\Phi$ and solutions exist with $\Phi$ constant. 
\item Constant $\tau$  and $k\neq 0$ which means the 6D fluxes come from the Stuckelberg potential.
\item Non-trivial profile for $\tau$ as  relevant for F-theory.
\end{enumerate}

We will concentrate here on scenario 1 which is the simplest to illustrate our results. We will dedicate a follow-up article to the detailed study of scenarios 2 and 3 which 
are more general and potentially more realistic.

\subsection{Ans\"atze and new asymptotic behaviour}

We repeat the arguments of \S\ref{Sec:OldSchool} to construct solutions to eqs.~\pref{eq:RMNeq} that are cylindrically symmetric in the extra dimensions and maximally symmetric in 4D.  For ease of reference, the equations to be solved are the Maxwell equation
\be\label{eq:Maxeqs2}
  \nabla_\ssM\Bigl( e^{-\varphi} F^{\ssM\ssN} \Bigr) = 0 \,,
\ee
and the scalar field equations
\be
\label{eq:phichieqs2}
  \Box \chi  + 4 \tilde{V} e^{\varphi - 2 \chi} = 0 \qquad \hbox{and} \qquad
 \Box\varphi + \tfrac14 \, \kappa^2 e^{-\varphi} F^{\ssM\ssN}F_{\ssM\ssN} - \tilde{V} e^{\varphi - 2 \chi} = 0 \,,
\ee
where $\kappa$ is the 6D Planck scale and $\tilde V$ is an independent dimensionful. Here $\varphi$ is the tensor-multiplet scalar and $\chi$ is the volume modulus of the Calabi Yau space. The 6D trace-reversed Einstein equation similarly is
\be 
\label{eq:RMNeq2}
R_{\ssM\ssN} +  \kappa^2 e^{-\varphi} F_{\ssM\ssP} {F^\ssP}_{\ssN} + \partial_\ssM\varphi  \, \partial_\ssN \varphi + \frac{1}{2} \partial_\ssM \chi \, \partial_\ssN \chi + \frac{1}{2} \, g_{\ssM\ssN} \Box_6 \varphi  = 0\,.
\ee

These agree with \pref{SSFEvarphi} through \pref{SSFEinstein} if evaluated at $\chi = 0$ for a particular choice of $\tilde V$, but we cannot simply use the solutions from \S\ref{Sec:OldSchool} because $\chi = 0$ is not a solution to the first of eqs.~\pref{eq:phichieqs2}. We instead use the same approach taken in \S\ref{Sec:OldSchool} to reconstruct new solutions. Notice that for 4D maximally symmetric solutions the 4D part of \pref{eq:RMNeq2} again relates the 4D curvature to a total derivative of $\varphi$, implying that the sign of the 4D curvature can be computed using only the near-source asymptotic form for $\partial_r \varphi$, along the lines of eq.~\pref{R4vsSurfTerms} \cite{Aghababaie:2003ar, Burgess:2011rv, Gautason:2013zw}.  

To this end we again follow \cite{Tolley:2005nu} and seek solutions with the 6D metric ansatz
\be \label{metricform1}
\exd s^2 =\hat g_{\mu\nu} \exd x^\mu \exd x^\nu + a^2 \exd\theta^2+a^2W^8 \exd\eta^2
\ee
where the coordinates $(\eta,\varphi)$ parametrise the compact two dimensions and $\hat g_{\mu\nu} = W^2 g_{\mu\nu}$ where $g_{\mu\nu}$ is the maximally symmetric 4D metric with curvature $R_{4}= - 3\zeta H^2$ and $\zeta = \pm 1$ or zero (with $\zeta = +1$ corresponding to de Sitter space). 2D axial symmetry requires $W = W(\eta)$, $a = a(\eta)$, $\varphi = \varphi(\eta)$ and $\chi = \chi(\eta)$ are all functions only of $\eta$. We expect an effective 6D regime to emerge when there is a hierarchy of size between the volume $\cV$ of the extra dimensions between 6 and 10 and the volume $a^2 W^4$ of the compact dimesions appearing within \pref{metricform1}. 

Under these assumptions the Maxwell equation implies $(e^{-\varphi}F_{\eta\theta}/a^2)'=0$ and so integrates to give
\be
F_{\eta\theta}= \hat Q \, a^2 e^\varphi
\ee
as before, with $\hat Q$ an integration constant. With this choice $F^{\ssM\ssN}F_{\ssM\ssN}=2 \hat Q^2 e^{2\varphi}/W^8$. The scalar field equations then are
\be \label{newscalareqs}
 \chi''= - 4\tilde V a^2 W^8 e^{\varphi-2\chi} \qquad \hbox{and} \qquad
 \varphi''=\tilde V a^2 W^8 e^{\varphi-2\chi}-\tfrac12 Q^2a^2 e^\varphi
\ee
where $Q := \kappa \hat Q$. In particular $\tilde V > 0$ implies $\chi'$ is a monotonically decreasing function of $\eta$. 

The $(\mu\nu)$ component of the Einstein equations are unchanged from \S\ref{Sec:OldSchool},
\be \label{newmunu}
 \Bigl( \ln W + \tfrac12 \varphi \Bigr)'' =  3 \zeta H^2a^2W^6 \,,
\ee
as is also the $\theta\theta$ component
\be \label{newthetatheta}
  \Bigl( \ln a + \tfrac12 \varphi \Bigr)'' = - Q^2a^2e^\varphi \,.
\ee
The $\eta\eta$ component can be expressed as the constraint 
\be \label{newconstraint}
  \varphi'^2+\tfrac{1}{2}\chi'^2-8(\ln W)'(\ln a)'-12[(\ln W)']^2  + a^2 e^\varphi\left( Q^2-2\tilde V W^8 e^{-2\chi}+12\zeta H^2W^6e^{-\varphi}\right) = 0 \,.
\ee

It can sometimes be useful to follow \cite{Tolley:2005nu} and notice that these ordinary differential equations can be regarded as arising from the `Lagrangian' 
\be
L := \Bigl[\varphi'^2 + \tfrac{1}{2}\chi'^2-8(\ln W)'(\ln a)'-12[(\ln W)']^2  \Bigr] N^{-1}-Na^2 e^\varphi\left( Q^2-2\tilde V W^8 e^{-2\chi}+12\zeta H^2W^6e^{-\varphi}\right)
\ee
because variations with respect to $\varphi$, $\chi$, $W$ and $a$ reproduce equations \pref{newscalareqs}, \pref{newmunu} and \pref{newthetatheta} after setting $N = 1$ while varying respect to $N$ gives the constraint \pref{newconstraint} (again after setting $N=1$).

\subsubsection{New asymptotic behaviour}

It is tempting to think that the similarity of these equations to those of \S\ref{Sec:OldSchool} will ensure that they will share the same near-source asymptotic form as was found in \pref{asymptoticpowers}, but the presence of the $\chi$ field makes this not quite true. 

To see why, it is useful to follow \cite{Tolley:2005nu} and change variables to decouple as many of the equations as possible. Defining new variables $\cX$, $\cY$ and $\cZ$ by
\be
   \varphi=\frac{1}{2}\left(\cX-\cY-2\cZ\right),\qquad \ln W=\frac{1}{4}\left( \cY -\cX\right),
   \qquad \ln a=\frac{1}{4}\left(3\cX + \cY +2\cZ\right)
\ee
and further
\be
X =  \cX + \ln Q \,, \qquad
Y =  \cY +\ln(2\tilde V) \quad \hbox{and} \quad
Z  =   \cZ +\ln\left(\frac{6H^2}{\tilde V}\right) \,,
\ee
allows the Lagrangian $L$ to be simplified to
\be
L= \Bigl[ \left(X'\right)^2-\left( Y'\right)^2+\left(Z'\right)^2+\tfrac{1}{2}\left(\chi'\right)^2 \Bigr] N^{-1} - N \Bigl[ e^{2X}- e^{2Y-2\chi} + \zeta e^{2Y+Z} \Bigr] \,,
\ee
leading to the simpler equations
\be \label{XYZeqns}
 X''+e^{2X}  =   0 \,, \quad \chi''+2e^{2Y-2\chi}  =   0 \,, \quad Z''+\tfrac{1}{2} \zeta e^{2Y+Z}=0  
 \quad \hbox{and} \quad
Y''+e^{2Y-2\chi}-\zeta e^{2Y+Z} =   0  \,.
\ee

The $X$ equation decouples from the others and can be integrated to give 
\be \label{Xsoln}
  X = \ln \left[ \frac{\lambda_1}{\cosh[\lambda_1(\eta -\eta_1) ]} \right] \,,
\ee
where $\lambda_1$ and $\eta_1$ are integration constants. The other three equations can be simplified by noticing $4Z''+2Y''-\chi''=0$ which implies  
\be
\chi = 4Z+2Y + b\eta+c
\ee
for integration constants $b$ and $c$. The remaining two equations depend only on $Y$ and $Z$ and can in principle be solved numerically. They are usefully rewritten by defining
\be
A :=\frac{1}{2}\left(2Y+Z\right) \qquad \hbox{and} \qquad B := Y-\chi=-(Y+4Z+b\eta +c)
\ee
because then the remaining equations for $Y$ and $Z$ become:
\be \label{ABeqs}
B''  =  \zeta e^{2A}+ e^{2B} \qquad \hbox{and} \qquad
A''  =  \tfrac{3}{4}\zeta  e^{2A}- e^{2B} \,.
\ee

Now comes the main point: the singularities of the solutions of these equations can differ significantly from those described in \S\ref{Sec:OldSchool} when $\chi$ is set to zero. To see why, it suffices to consider the case of 4D Minkowski solutions, for which $\zeta =0$. In this case the $B$ equation completely decouples to become
\be \label{FlatBeq}
B''=e^{2B}  \qquad \hbox{(4D flat solutions)}
\ee
and can be directly integrated, with solution
\be \label{Bsoln}
  B=\ln \left[\frac{\lambda_q}{\sinh[\lambda_q(\eta-\eta_0)]}\right]  \,,
\ee
where the difference between this and \pref{Xsoln} has its roots in the difference in the sign of the exponential in \pref{FlatBeq} relative to  the $X$ equation in \pref{XYZeqns}.

What is significant about \pref{Bsoln} is the appearance of a new singularity at $\eta = \eta_0$ in addition to the `old' singularities at $\eta \rightarrow \pm \infty$ familiar from \S\ref{Sec:OldSchool}. Even if we start at $\eta \to - \infty$ using one of the asymptotic Kasner-type solutions given in \pref{asymptoticpowers} with $\chi = \chi' = 0$ the evolution of $\chi$ ensures that the solution does {\it not} evolve towards another Kasner-type solutions at $\eta \to + \infty$ because a new type of singularity instead intervenes as $\eta \to \eta_0$. As we shall see, a new class of asymptotic behaviour emerges because the solutions evolve towards configurations for which the non-derivative terms on the right-hand sides of eqs.~\pref{newscalareqs}, \pref{newmunu} and \pref{newthetatheta} are not negligible (that is, the inequalities \pref{inequalities} assumed when developing the approximate solutions \pref{asymptoticpowers} break down). 

We wish to identify the asymptotic behaviour of the fields near this new singularity and see what these imply for the properties of any sources located at the singularity, though before doing so rewrite the equations in terms of the proper distance along the extra dimensions.

\subsection{Solutions as functions of proper distance}

Asymptotic solutions are easier to obtain (and the nature of the sources to which they point are easier to identify) if we change coordinates from $\eta$ to proper distance $r$ satisfying $\exd r=aW^4 \exd\eta $. This section therefore sets up and solves (both asymptotically and numerically) the field equations using proper distance.

It is also convenient to introduce new logarithmic variables $a =: e^{\Omega}$ and $W =: e^{\Gamma}$ so that the metric becomes
\be
\exd s^2 
=e^{2\Gamma(r)} g_{\mu\nu} \exd x^\mu \exd x^\nu+ e^{2\Omega(r)} \exd\theta^2+\exd r^2 
\ee
and proper distance becomes $\exd r = e^{\Omega + 4 \Gamma} \exd\eta$.   

Denoting $\exd/\exd \eta$ with primes and $\exd/\exd r$ with over-dots, for any function $J[\eta(r)]$ we have
\be
   J' = e^{\Omega+4\Gamma} \dot J \quad \hbox{and} \quad J'' = e^{2\Omega+8\Gamma} \Bigl[\ddot J+\left(\dot\Omega+4\dot\Gamma\right)\dot J \Bigr]\,,
\ee 
and so the field equations \pref{newscalareqs} through \pref{newthetatheta} become
\bea  \label{propreqs}
\ddot\varphi +\left(\dot\Omega+4\dot \Gamma \right)\dot\varphi & = & \tilde Ve^{\varphi -2\chi}-\tfrac12 Q^2 e^{\varphi - 8\Gamma}\nonumber\\
\ddot\chi+\left(\dot\Omega+4\dot \Gamma \right)\dot\chi & = &  -4\tilde V e^{\varphi-2\chi}\nonumber\\
\ddot\Gamma+\left(\dot\Omega+4\dot \Gamma \right)\dot\Gamma & = &3\zeta H^2 e^{-2\Gamma}-\tfrac12 \left[\ddot\varphi+\left( \dot\Omega+4\dot \Gamma \right)\dot\varphi\right]\\ 
\ddot\Omega+\left(\dot\Omega+4\dot \Gamma \right)\dot\Omega & = &-Q^2 e^{\varphi-8\Gamma} -\tfrac12 \left[\ddot\varphi+\left(\dot\Omega+4\dot \Gamma \right)\dot\varphi\right] \,.\nonumber
\eea
These imply $Q$, $\tilde V$ and $H$ drop out of the combination
\be \label{DerivOnly}
\ddot\chi +3\ddot \varphi -2\ddot\Omega=- \left(\dot\Omega+4\dot\Gamma\right)\left(\dot\chi+3\dot\varphi-2\dot\Omega\right) \,,
\ee
and allow the total derivatives \pref{newmunu} and \pref{newthetatheta} to be written
\be \label{newtotderivsr}
 \frac{\exd}{\exd r} \Bigl[e^{\Omega + 4\Gamma} \Bigl(\dot \Gamma + \tfrac12 \dot\varphi \Bigr) \Bigr] =  3 \zeta H^2 e^{\Omega + 2 \Gamma}
 \quad \hbox{and} \quad
  \frac{\exd}{\exd r} \Bigl[e^{\Omega + 4\Gamma} \Bigl(\dot \Omega + \tfrac12 \dot\varphi \Bigr) \Bigr] =  - Q^2 e^{\varphi +\Omega- 4 \Gamma} \,.
\ee
Finally, the constraint \pref{newconstraint} becomes 
\be \label{proprconstraint}
 \dot\varphi^2+\tfrac{1}{2}\dot\chi^2 -8\dot\Omega\dot\Gamma-12\dot\Gamma^2+ 12\zeta H^2e^{-2\Gamma}+Q^2e^{\varphi-8\Gamma}-2\tilde V  e^{\varphi - 2 \chi} =0\,.
\ee

\subsubsection{Kasner-type solutions}

We seek solutions to these equations with scaling behaviour close to $r=0$ (the singularity's location) of the form\footnote{Since $a$ has dimensions of length $a_0 = a(r_0) = e^{\Omega_0} = \ell$ defines a length scale in this asymptotic expression.}
\be \label{fieldasymforms}
 \varphi = \varphi_0 + q \ln \left(\frac{r}{r_0}\right) , \quad
 \chi  = \chi_0 + s \ln \left(\frac{r}{r_0}\right)  , \quad
 \Gamma  = \Gamma_0 +  w \ln \left(\frac{r}{r_0}\right) \quad \hbox{and} \quad
 \Omega  =  \Omega_0 + \alpha \ln \left(\frac{r}{r_0}\right)  ,
\ee
up to neglected terms that go to zero when $r \to 0$. The last of these is equivalent to
\be
a(r)= a_0\left( \frac{r}{r_0} \right)^\alpha \Bigl[ 1+{\mathcal O}(r)\Bigr], \qquad \text{for}\qquad a(r)\propto e^\Omega
\ee
which emphasizes that for this case there is a length scale $a_0$ hidden within $\Omega_0$. For this leading behaviour $\dot\varphi = q/r$ and $\ddot \varphi = -q/r^2$ and so on. The $1/r^2$ terms in eqs.~\pref{propreqs} and \pref{proprconstraint} dominate as $r \to 0$ provided 
\be \label{KIneq}
   q > 2(s-1) \qquad \hbox{and} \qquad q > 2(4w-1) \qquad \hbox{and (if $H \neq 0$) } \; w < 1 \,, 
\ee
and when this is true the equations imply the Kasner-like conditions
\be \label{KasnerLinear2}
  \alpha + 4 w = 1  
\ee
and
\be \label{KasnerQuad2}
  q^2 + \tfrac12 s^2 - 8\alpha w - 12 w^2 = q^2 + \tfrac12 s^2 - 8w + 20 w^2 = 0 
\ee
where the first equality uses \pref{KasnerLinear2} to eliminate $\alpha$. 

Eqs.~\pref{KasnerLinear2} and \pref{KasnerQuad2} determine $s$ and $\alpha$ in terms of $w$ and $q$, and the inequalities put constraints on the allowed region in the $(q,w)$ plane. Positivity of $q^2 + \frac12 s^2$ implies $0 \leq w \leq \frac25$, and for any real $w$ we have $q^2 + \frac12 s^2 \leq \frac45$ with the maximum obtained when $w = \frac15$. Requiring circles of radius $r$ to have circumferences that shrink as $r \to 0$ further requires $\alpha > 0$ and this implies $w < \frac14$. Notice that these automatically fall into the $w< 1$ region required if $H\neq 0$. There is plenty of room to find nonzero values for $q$ and $w$ for which all inequalities in \pref{KIneq} are satisfied, and the resulting solutions generalize the Kasner-like solutions of \S\ref{Sec:OldSchool} to include a nontrivial profile $\chi(r)$. Of these the most trivial solutions have $s = q = 0$, corresponding to situations were the scalars $\varphi$ and $\chi$ do not diverge at the source locations. For these eqs.~\pref{KasnerLinear2} and \pref{KasnerQuad2} imply $w = 0$ and $\alpha = 1$, so the warp factor does not diverge or get driven to zero and the geometry has a conical singularity at $r = 0$. 

\subsubsection{New asymptotic solutions}

As noted above, the Kasner-like solutions are not sufficiently general to capture all of the singularities found when integrating \pref{ABeqs} such as found in particular in \pref{Bsoln}. To capture the asymptotic form for these requires no longer assuming that the derivative terms dominate, so we ask under which conditions the non-derivative terms compete with the derivative terms for small $r$. That is, we ask when the non-derivative terms can also scale as $1/r^2$ rather than being subdominant. 

The leading power of $r$ in eqs.~\pref{propreqs} then becomes 
\bea  \label{propreqs2}
 \frac{q}{r^2} \Bigl(-1 + \alpha + 4w \Bigr) & = & \tilde Ve^{\varphi_0 -2\chi_0} \left( \frac{r_0}{r} \right)^{2s-q} -\tfrac12 Q^2 e^{\varphi_0 - 8\Gamma_0} \left( \frac{r_0}{r} \right)^{8w-q} \nonumber\\
 \frac{s}{r^2} \Bigl(-1 + \alpha + 4w \Bigr)  & = &  -4\tilde V e^{\varphi_0-2\chi_0} \left( \frac{r_0}{r} \right)^{2s-q}\nonumber\\
 \frac{w+\frac12 q}{r^2} \Bigl(-1 + \alpha + 4w \Bigr)  & = &3\zeta H^2 e^{-2\Gamma_0}\left( \frac{r_0}{r} \right)^{2w} \\ 
 \frac{\alpha+\frac12 q}{r^2} \Bigl(-1 + \alpha + 4w \Bigr) & = &-Q^2 e^{\varphi_0 - 8\Gamma_0} \left( \frac{r_0}{r} \right)^{8w-q}  \,.\nonumber
\eea
It is clear that new solutions can exist when the inequalities \pref{KIneq} are saturated, but if they do they cannot satisfy the linear Kasner condition $\alpha + 4w = 1$. The linear combination \pref{DerivOnly} is particularly simple and implies
\be \label{DerivOnly2}
  (s+3q - 2\alpha)(-1 + \alpha + 4w) \frac{1}{r^2} = 0 \,.
\ee
The constraint \pref{proprconstraint} similarly generalizes to
\be \label{proprconstraint2}
 \frac{q^2 + \tfrac12 s^2 - 8\alpha w - 12 w^2}{r^2}   + 12\zeta H^2e^{-2\Gamma_0}\left( \frac{r_0}{r} \right)^{2w}+Q^2e^{\varphi_0 - 8\Gamma_0} \left( \frac{r_0}{r} \right)^{8w-q}-2\tilde V  e^{\varphi_0 -2\chi_0} \left( \frac{r_0}{r} \right)^{2s-q}=0\,.
\ee

Some or all of the non-derivative terms can compete with the $1/r^2$ terms if some or all of the following conditions hold
\be \label{Marginal}
 q = 2(s-1) \,, \qquad q = 2(4w-1)  \qquad \hbox{and (if $H \neq 0$)} \; w = 1 \,.
\ee 
When these hold the sum of the coefficients of $1/r^2$ in the corresponding equations in \pref{propreqs2} and \pref{proprconstraint2} must also vanish. Consider, for instance, the most restrictive case where all three conditions in \pref{Marginal} hold. These together with \pref{DerivOnly2} then completely determine the powers, giving $w = 1$, $q = 6$, $s = 4$ and  $\alpha = 11$. The last three equations of \pref{propreqs2} can then be used to solve for the three quantities $\tilde V \, e^{\varphi_0 - 2\chi_0} r_0^{2}$, $Q^2 e^{\varphi_0-8\Gamma_0} r_0^{2}$ and $\zeta H^2 e^{-2\Gamma_0} r_0^{2}$. This leaves the constraint  \pref{proprconstraint2} but as is easily checked this is automatically satisfied once the other conditions are (as might have been expected due to the Bianchi identity). The problem with this solution is that the last of eqs.~\pref{propreqs2} implies $Q^2 e^{\varphi_0-8\Gamma_0} r_0^{2} = -(14)^2$ is negative. A similar problem arises if all terms are required to scale like $r^{-2}$ when $H = 0$, which requires $w = \frac15$, $q = - \frac25$, $s = \frac45$ and $\alpha = - \frac15$, and again $Q^2$ must be negative. 

An existence proof that a solution of this new type exists is the case where the $\tilde V$ term scale as $1/r^2$ but the $Q^2$ and $H^2$ terms are subdominant. In this case \pref{propreqs2} and \pref{DerivOnly2} imply 
\be \label{nonKasnerpowers}
   w = \tfrac19 \,, \quad q = - \tfrac29 \,, \quad s = \tfrac89 \,, \quad \alpha =  \tfrac19 \quad \hbox{and} \quad
   \tilde V  e^{\varphi_0 -2\chi_0} r_0^2 = \tfrac{8}{81}  \,,
\ee
which also satisfies $8w-q = \frac{10}3 < 2$ and $w < 1$ (as required for the subdominance of the $Q^2$ and $H^2$ terms). This is the class of solutions to which we find our numerical solutions typically evolve. 

\subsubsection{Numerical 4D de Sitter solutions}

We have numerically integrated the evolution equations \pref{propreqs} forward in $r$ starting from initial conditions at $r = r_{ic}$ for the fields and their derivatives that are chosen to satisfy the constraint \pref{proprconstraint}. We verify numerically that the constraint remains satisfied for other values of $r$ as a check on calculations. 

In practice we choose $r_{ic}$ to be very small and choose initial values consistent with an asymptotic solution of the Kasner-like form \pref{fieldasymforms} that diverges at $r = 0$. The Kasner parameters $\alpha$ and $q$ are specified and then $s$ and $w$ are determined from the Kasner constraints \pref{KasnerLinear2} and \pref{KasnerQuad2}. The equations are then integrated numerically to larger values of $r$ until they again diverge, at a proper distance denoted by $r_e$. We numerically compare the solution's asymptotic form near this second singularity and verify that it also satisfies the power-law form \pref{fieldasymforms}. This comparison also reveals the powers $q$, $s$, $\alpha$ and $\omega$ for this second singularity. For the solution displayed these turn out to be given by \pref{nonKasnerpowers}.

\begin{figure}[h]
    \centering
    \begin{subfigure}[b]{0.45\linewidth}
        \includegraphics[width=\linewidth]{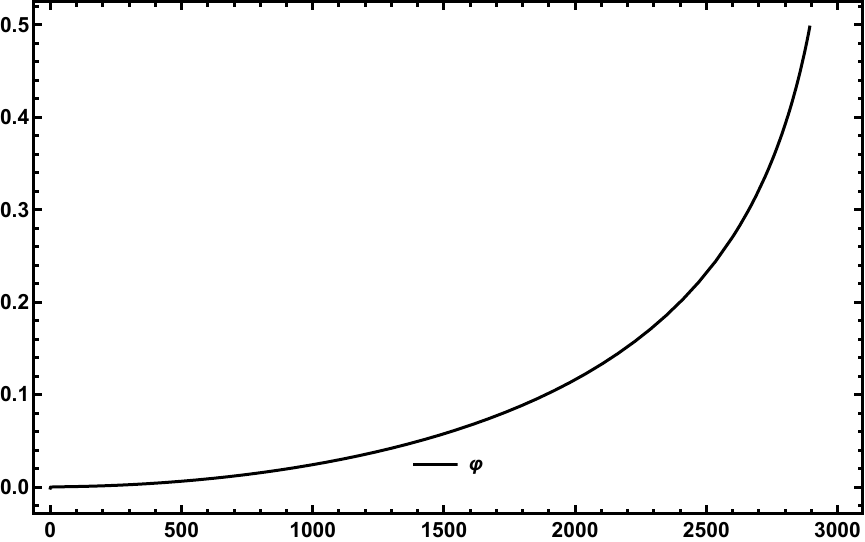} 
        \caption{The function $\varphi(r)$ vs $r$}
        \label{fig:PhiP}
    \end{subfigure}
    \hfill 
    \begin{subfigure}[b]{0.45\linewidth}
        \includegraphics[width=\linewidth]{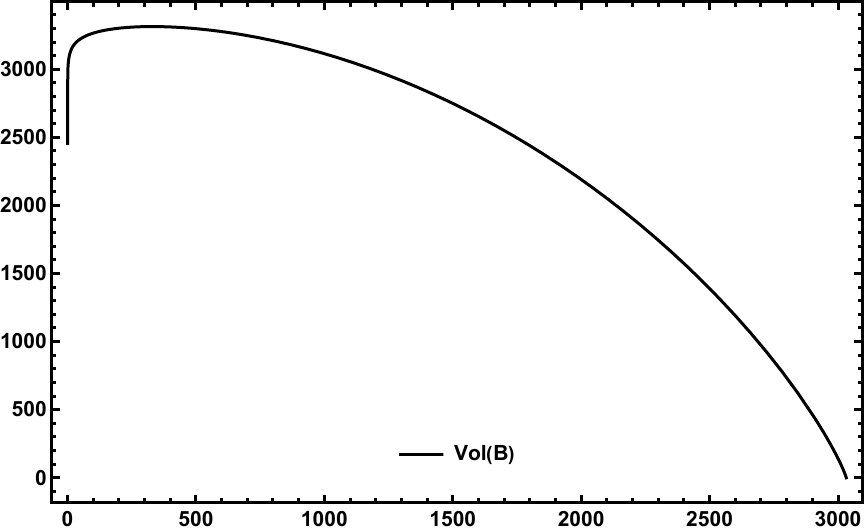} 
        \caption{The function $\cV  = e^{\chi(r)}$ vs $r$.}
        \label{fig:ChiP}
    \end{subfigure}
    \vspace{2mm}
    \begin{subfigure}[h]{0.45\linewidth}
        \includegraphics[width=\linewidth]{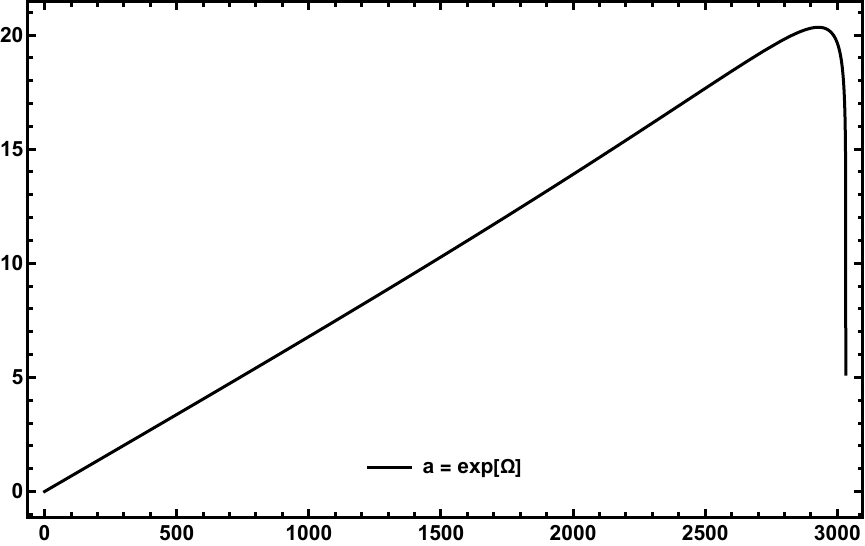} 
        \caption{The function $a(r) = e^{\Omega(r)}$ vs $r$.}
        \label{fig:OmegaP}
    \end{subfigure}
    \hfill  
    \begin{subfigure}[h]{0.45\linewidth}
        \includegraphics[width=\linewidth]{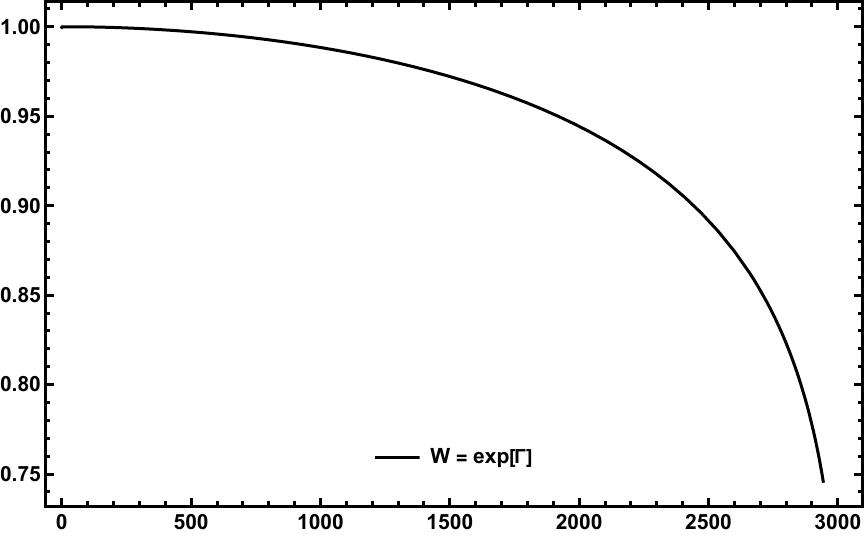} 
        \caption{The function $W(r) = e^{\Gamma(r)}$ vs $r$.}
        \label{fig:GammaP}
    \end{subfigure}
    
   \caption{\small Solutions to eqs.~\eqref{propreqs} {\it vs} proper distance with initial conditions chosen near $r=0$ consistent with 4D de Sitter geometry and a Kasner asymptotic form with powers $q = 0.0001$, $\alpha = 0.9999$, $s = 0.0200$ and $w = 0.000024$. The numerical prefactors satisfy the constraint equation with coefficients $\Omega_0 = -5$, $\chi_0 = 8$ and $H=10^{-5}$ in units where $\tilde V = 1$. The second singularity is arises at $r = 3032$ with an asymptotic power-law form with the non-Kasner powers given in \pref{nonKasnerpowers}. 
    }
    \label{fig:fourplotsP}
\end{figure}

Figure \ref{fig:fourplotsP} show the results of numerical evolution obtained in this way with initial conditions specified at $r = r_{ic} = 10^{-6}$, where the numerics use units\footnote{As noted below eq.~\pref{eq:Scalars6DF} the choice $\mfg^2/\kappa \ll 1$ ensures the energy unit specified by $\tilde V = 1$ is much smaller than 6D Planck size.} for which $\tilde V = 1$. The initial conditions correspond to an asymptotic power-law form with Kasner powers
\begin{equation}
  \alpha_0 = 0.9999 \, \qquad q_0 = 0.0001 \,, \qquad s_0 \simeq 0.0200 \,, \qquad w_0 = 0.000025 
\end{equation}
chosen to satisfy the constraints \pref{KasnerLinear2} and \pref{KasnerQuad2} and to be not far from the trivial solution (for which $s=q=w=0$ and $\alpha = 1$). The solution also assumes $\Omega_0 = -5$, $\chi_0 = +8$ and $\Gamma_0 = \varphi_0 = 0$ in the assumed small-$r$ asymptotic form \pref{fieldasymforms}. We use the constraint \pref{proprconstraint} to generate the rest of the initial conditions assuming the 4D maximally symmetric dimensions form a de Sitter space (and so $\zeta = +1$) for which we choose the Hubble scale $H = 10^{-5}$. 

The field equations are then integrated numerically towards increasing $r$ until a second singularity is encountered, which for the solution displayed occurs at a proper distance $r_e = 3032$ from the first Kasner singularity. The four panels of Fig.~\ref{fig:fourplotsP} respectively plot $\varphi(r)$, the volume of the transverse compact 4D space $B_4$ (in string units) $\cV = e^{\chi(r)}$ together with the 6D metric functions $a(r) = e^{\Omega(r)}$ and $W(r) = e^{\Gamma(r)}$ obtained in this way. All are plotted as a function of proper distance away from the Kasner-type source.

\begin{figure}[h]
    \centering
    \begin{subfigure}[h]{0.45\linewidth}
        \includegraphics[width=\linewidth]{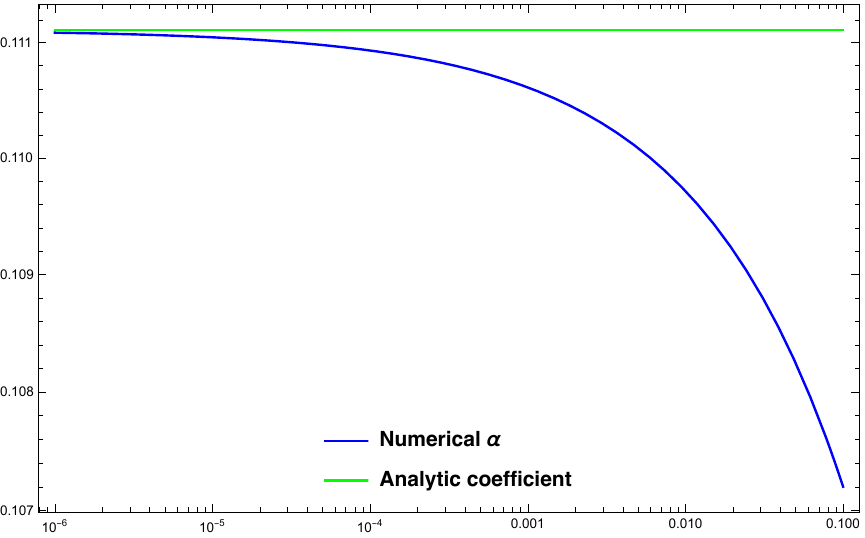}  
        \caption{$r\, \partial \Omega/\partial r$ {\it vs} $\alpha_e = \frac19$}
        \label{fig:PhiComp}
    \end{subfigure}
    \hfill 
    \begin{subfigure}[h]{0.45\linewidth}
        \includegraphics[width=\linewidth]{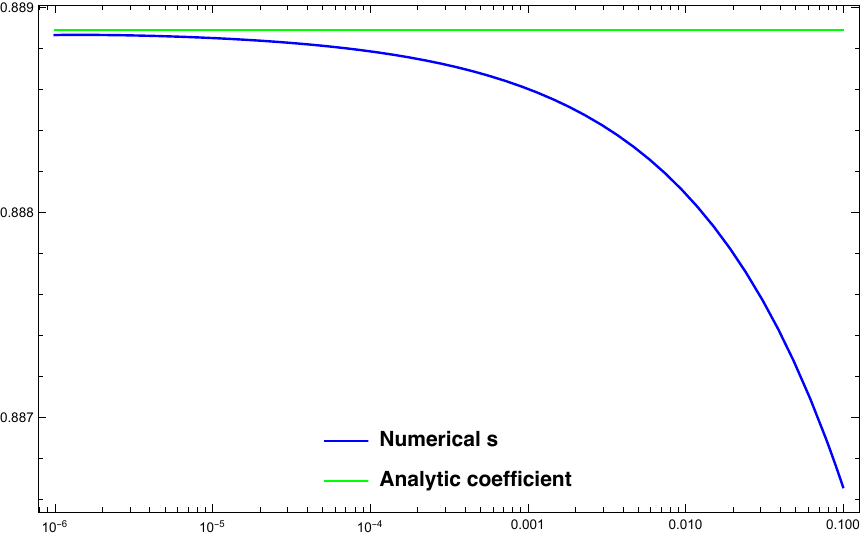}  
        \caption{$r\, \partial \chi/\partial r$ {\it vs} $s_e = \frac89$}
        \label{fig:PhiCompZoom}
    \end{subfigure}
        
    \caption{\small Plot of $r \,\partial \Omega/\partial r$ and $r \,\partial \chi/\partial r$ vs $r_e - r$ for the solution in Fig.~\ref{fig:fourplotsP}, which should be $r$-independent in the scaling regime of \pref{fieldasymforms}. Also shown is the prediction for $\alpha_e$ and $s_e$ obtained from the non-Kasner power-law \pref{nonKasnerpowers}.}
    \label{fig:varphiPowerTransition}
\end{figure}

The asymptotic form of the solution near this second singularity is found numerically to agree with the power-law form given in \pref{fieldasymforms}. This comparison also reveals what the powers $q_e$, $s_e$, $w_e$ and $\alpha_e$ are for this asymptotic solution. This can be seen, for instance, in  Fig.~\pref{fig:varphiPowerTransition} which computes $r \, \partial \Omega/\partial r$ and $r \, \partial \chi/\partial r$ as $r \to r_e$ (a quantity that should be $r$-independent -- and equal to $\alpha_e$ and $s_e$ respectively -- if the asymptotic solution \pref{fieldasymforms} applies). These comparisons for the solution displayed reveal that the powers agree with those predicted by \pref{nonKasnerpowers} once $r_e - r$ is smaller than $10^{-5}$. 

We can check {\it ex post facto} that this solution lies within the domain of validity of the derivation of the 6D field equations. 
\begin{itemize}
\item {\it Weak string coupling:} is ensured by choosing the field $\tau$ appropriately, and $\tau$ is constant in our solutions with a value not fixed by the field equations. We are therefore free to choose $\tau$ to be in the weak-coupling regime and once this is done it follows that the string length $\ell_s$ (defined by $\alpha'$) satisfies $\ell_s = \ell_{10}/\lambda$ where $\ell_{10}$ is the 10 Planck length (defined from $\kappa_{10}$) and $\lambda \ll 1$.

\item {\it 10D supergravity regime:} Use of the higher-dimensional supergravity equations requires the four compact dimensions $B_4$ transverse to the 6D theory to be large in string units. In our conventions this requires $\cV = e^{\chi} \gg 1$, since $\cV$ is defined to be the volume of $B_4$ measured in string units: $\cV = \ell_\ssB^4/\ell_s^4$. This is satisfied by the top-right panel of Fig.~\ref{fig:fourplotsP} except in the immediate vicinity of the right-hand (non-Kasner) singularity. 

\item {\it Effective 6D limit:} requires the length scale $\ell_\ssB$ setting the linear size of the four transverse dimensions, $B_4$, must be larger than the scales defining the 6D theory. (This hierarchy is required for solutions of the higher dimensional field equations to be well-approximated by the 6D ones found explicitly here.) On one had the lower-left panel of Fig.~\ref{fig:fourplotsP} shows that the proper distance $r_e \sim 3000$ and the circumference $2\pi a(r)$ of circles at fixed $r$ are both large compared with unity and therefore also compared with the 6D Planck length $\ell_6$ set by the value $\kappa$.  (As discussed below eq.~\pref{eq:Scalars6DF} choosing $\mfg^2 \ll \kappa$ and $\tilde V = 1$ ensures $\ell_6 \ll 1$. ) On the other hand we have also seen that $\ell_\ssB = \cV^{1/4} \ell_s = \cV^{1/4} \ell_{10}/\lambda$ where weak string coupling says $\lambda \ll 1$. But compactification predicts the 6D and 10D Planck lengths are in turn related by $\ell_{10} \sim \ell_6 \, \cV^{-1/4} \ll \ell_6 \ll 1$ so we have $L_{\ssB} \sim  \ell_{10}/\lambda$. For instance if we choose $\tau$ so that $\lambda \sim \cV^{-1/4} \ll 1$ then $\ell_\ssB \sim \ell_6$ is much smaller than the compact two dimension in 6D (except very close to the right-hand (non-Kasner) singularity. 

\item {\it Semiclassical methods in 6D:} As usual for gravity (see {\it e.g.}~\cite{EFTBook}) semiclassical methods are controlled by powers of $\ell_6$ divided by the curvature radii which in the 6D theory are set by extra-dimensional size and by the 4D Hubble length $H^{-1}$. This is under control for our solutions because the curvature radii are both much larger than unity but $\ell_6$ must be smaller than unity.

\end{itemize} 
 
We see that the various small control parameters in the analysis can be kept controllably small, except perhaps in the immediate vicinity of the right-hand non-Kasner singularity.

\subsection{Interpretation of the singularities}

The singularities in the bulk solutions can again be matched to the action of the gravitating source whose back-reaction is responsible for the singular behaviour along the lines described in \S\ref{Sec:OldSchool}. We again assume a source action containing the fewest derivatives, as in \pref{BraneAction}, leading for unbent sources to
\be \label{BraneAction2}
  S_b   = - \int \exd^4 x \sqrt{- g} \; W^4_b L_b(\varphi, \chi) = - \int \exd^4 x \sqrt{- g} \; T_b (\varphi , \chi)\, .
\ee

The matching conditions that follow when \pref{BraneAction2} is combined with eqs.~\pref{GaugePPEFT} through \pref{MetricPPEFT}, can be specialized to the field equations \pref{eq:phichieqs2} and \pref{eq:RMNeq2} and simplified by cancelling a common factor of $W^4(\epsilon)$. This leads to the following near-source matching relation for the scalars  
\be \label{scalarmatchingNS2phicombo}
   \Bigl[  a\, \partial_r \, \varphi \Bigr]_{r=\epsilon}   =  \left( \frac{\partial \cL_b}{\partial \varphi} \right)_{r = \epsilon}  
\qquad \hbox{and} \qquad 
    \Bigl[  a   \, \partial_r \, \chi \Bigr]_{r=\epsilon}   = \left( \frac{\partial \cL_b}{\partial\chi} \right) _{r=\epsilon}\,,
\ee
where $\cL_b = \kappa^2 L_b/(2\pi)$ is a dimensionless measure of the strength of the gravitational back-reaction of the source. These equations show precisely how the singularity in a bulk scalar at the source location is controlled by the strength with which it couples to the source action. The metric matching conditions similarly are
\be  \label{munumatchingW2}
   1- \left[a \left( 3  \frac{ \partial_r W}{W}  + \frac{\partial_r a}{a} \right)  \right]_{r=\epsilon}  = \Bigl( \cL_b \Bigr)_{r= \epsilon}  \qquad \hbox{and} \qquad
  \Bigl[ a  \, \partial_r \, W  \Bigr]_{r = \epsilon}  = \Bigl(   \tilde \cU_b \Bigr)_{r =\epsilon} \,,
\ee
where $\tilde \cU_b := \kappa^2 \tilde U_b/(2\pi)$ with $\tilde U_b$ defined below eq.~\pref{thetathetamatchingW}. 

Just as in earlier sections, although the right-hand sides of the first three of these can be read off from \pref{BraneAction2} the same is not true for $\tilde \cU_b$. But this need not pose a problem because $(\partial_r W)_{r=\epsilon}$ can be read off in terms of the other derivatives and fields using the constraint \pref{proprconstraint}, and once this is done the second of eqs.~\pref{munumatchingW2} can be regarded as determining $\tilde \cU_b$. 

Further simplifying using the near-source asymptotic forms \pref{fieldasymforms} then allows \pref{scalarmatchingNS2phicombo} to be written
\be \label{scalarmatchingNS2phiw}
     q_b \left( \frac{a_b}{r_b} \right) \left( \frac{\epsilon}{r_b} \right)^{\alpha_b -1}     =  \left( \frac{\partial \cL_b}{\partial \varphi} \right)_{r = \epsilon}  
\qquad \hbox{and} \qquad 
   s_b \left( \frac{a_b}{r_b} \right) \left( \frac{\epsilon}{r_b} \right)^{\alpha_b -1}     = \left( \frac{\partial \cL_b}{\partial\chi} \right) _{r=\epsilon}\,,
\ee
where the arbitrary pivot scale $r_0$ is denoted $r_b$ to emphasize that it can differ from source to source while $a_b := a(r_b)$ and so on for the other fields. The metric conditions \pref{munumatchingW2} similarly become
\be \label{munumatchingW2w}
      1 -  ( 3 w_b   + \alpha_b) \left( \frac{a_b}{r_b} \right)\left( \frac{\epsilon}{r_b} \right)^{\alpha_b   -1}   = \Bigl( \cL_b \Bigr)_{r= \epsilon}  
\ee
and
\be \label{thetathetamatching2w}
  w_b \left( \frac{a_b}{r_b} \right) \left( \frac{\epsilon}{r_b} \right)^{\alpha_b  -1}  = \Bigl(  \tilde \cU_b \Bigr)_{r =\epsilon}\,.
\ee

We next consider two cases: weakly coupled sources for which $\cL_b = \kappa^2 L_b/(2\pi) \ll 1$ (and the same is true for its derivatives with respect to $\varphi$ and $\chi$) and generic sources for which these quantities need not be small. 

\subsubsection{Weakly gravitating (Kasner) sources}

For weakly coupled sources we assume $\cL_b$ and $\tilde \cU_b$ are both small, in which case \pref{scalarmatchingNS2phiw} and \pref{thetathetamatching2w} imply that $q_b$, $s_b$ and $w_b$ are small while \pref{munumatchingW2w} implies the same for $\alpha_b-1$. This puts them perturbatively close to the trivial Kasner-like solution for which $q=s=w=0$ and $\alpha = 1$. Dropping subdominant terms in $\cL_b$ and $\cU_b$ allows the neglect of $\alpha_b -1$ on the left-hand sides of the matching conditions, and so $(\epsilon/r_b)^{\alpha_b-1} \simeq 1$. Under these circumstances the matching conditions to simplify to
\be \label{shortmatch}
    q_b \left( \frac{a_b}{r_b} \right)   =  \left( \frac{\partial \cL_b}{\partial \varphi} \right)_{r = \epsilon} \,, \quad
     s_b \left( \frac{a_b}{r_b} \right)  = \left(\,  \frac{\partial \cL_b}{\partial\chi} \right) _{r=\epsilon} \,, \quad
     w_b \left( \frac{a_b}{r_b} \right)  = \Bigl(\tilde \cU_b \Bigr)_{r=\epsilon}  
\ee
and
\be \label{shortmatch2}
       1 - ( 3 w_b   + \alpha_b) \left( \frac{a_b}{r_b} \right) =  \Bigl( \cL_b \Bigr)_{r= \epsilon} \,.
\ee

In this Kasner-like limit non-derivative terms can be dropped as being subdominant in powers of $r$ in the near-source limit and so the constraint \pref{proprconstraint} can be written 
\bea \label{einsteinconstraintU2}
    0 &\simeq&  \left[ \dot\varphi^2+\tfrac{1}{2}\dot\chi^2 -8\dot\Omega\dot\Gamma-12\dot\Gamma^2\right]_b = q_b^2   + \tfrac12 s_b^2 - 8\alpha_b w_b - 12w_b^2  \nn\\
    &=& W_b^4 \Bigl[ - 8\, \tilde\cU_b \Bigl( 1 - \cL_b - 3 \tilde \cU_b \Bigr) - 12 \, \tilde \cU_b^2 + (\cL_{b\,,\varphi})^2+\tfrac12 (\cL_{b\,,\chi})^2 \Bigr]\,,
\eea
where the last line uses the matching conditions \pref{shortmatch} and \pref{shortmatch2} and subscripts $\varphi$ and $\chi$ denote differentiation with respect to the corresponding field. Once solved for $\tilde \cU_b$ this gives
\be \label{Ubsoln2}
 \tilde \cU_b = \tfrac13 \left[ (1 - \cL_b) - \sqrt{(1 - \cL_b)^2
 - \tfrac34 \, (\cL_{b\,,\varphi})^2 - \tfrac38 \, (\cL_{b\,,\chi})^2} \right]  
 \simeq  \tfrac18 \, (\cL_{b\,,\varphi})^2 + \tfrac{1}{16} \, (\cL_{b\,,\chi})^2 + \cdots\,,
\ee
where the approximate equality works to leading order in $\cL_b$. We see that $\tilde \cU_b$ vanishes if $\cL_b$ is independent of $\chi$ and $\varphi$ and once used in \pref{shortmatch} shows that the power $w_b$ is 
\be  \label{wissmall}
         w_b  \left( \frac{a_b}{r_b} \right)  \simeq   \frac18 \, \left( \frac{\partial \cL_b}{\partial \varphi} \right)^2_{r = \epsilon} + \frac{1}{16} \, \left(\,  \frac{\partial \cL_b}{\partial\chi} \right) ^2_{r=\epsilon}  + \cdots\,,
\ee
and so can be neglected because it is quadratic in $\cL_b$ and its derivatives.
 
Because \pref{wissmall} implies $w_b$ is second order in $\cL_b$ it can be neglected at leading order, in which case the Kasner relation \pref{KasnerLinear2} implies $\alpha_b = 1- 4w_b \simeq 1$ and \pref{shortmatch2} simplifies to
\be  \label{munumatchingW2zs}
      \frac{a_b}{r_b}    \simeq 1- \Bigl( \cL_b \Bigr)_{r= \epsilon}  \,.
\ee
Since the near-source metric has the form $\exd r^2 + a^2 \exd \theta^2 \simeq \exd r^2 + a_b^2 (r/r_b)^2 \exd \theta^2$ when $\alpha_b \simeq 1$ we see that the geometry has a conical defect $\delta_b$ whose size is related to $\cL_b$ in the standard way
\be \label{deltabvsLb}
   \delta_b = 2\pi \left( 1 -  \frac{a_b}{r_b} \right)  \simeq \kappa^2 L_b \,.
\ee 
In particular positive defect corresponds to positive tension. We see in this way that weak coupling leads to Kasner-like solutions that are perturbatively close to the trivial solution $w=s=q=0$ and $\alpha = 1$, and the sign of the tension is related to the sign of the bulk geometry's conical defect angle.

The sign of $L_b$ can be read off from the defect angle of the geometry using \pref{deltabvsLb}, at least for the Kasner-type singularities. To see how, denote by $\ell$ the length scale we set to unity when we take $\tilde V = 1$. Then the relation $a(r)/\ell = e^{\Omega(r)}$ together with the asymptotic forms $a(r) = a_b(r/r_b)^\alpha$ and $\Omega(r) = \Omega_0 + \alpha \log(r/\ell)$ shows that $\Omega_0$ is given by
\be
  \Omega_0 = \log \left( \frac{a_b}{\ell} \right) + \alpha \log \left( \frac{\ell}{r_b} \right) = \log \left( \frac{a_b}{r_b} \right) - (1 - \alpha) \log \left( \frac{\ell}{r_b} \right) \,.
\ee
But from \pref{deltabvsLb} we see that $L_b > 0$ requires $a_b < r_b$. Recall that $\ell$ is a macroscopic scale associated with the extra dimensions and $r_b$ is a microscopic scale associated with the source so we expect $\ell > r_b$. It follows that if $\alpha < 1$ (such as is true for a Kasner-like singularity) then $\Omega_0$ must be negative if $a_b < r_b$. In particular, if $1-\alpha \ll 1$ and $\Omega_0$ is negative and order unity (as is true for the numerical solution shown) then $L_b > 0$.  

In the special case where $\cL_b$ is independent of $\chi$ and $\varphi$ we see that $q_b = s_b = 0$ and the Kasner conditions then imply $w_b = 0$ and $\alpha_b = 1$. The matching conditions boil down to the usual relation between the conical defect angle and the source's tension $\cL_b = \kappa^2 L_b/2\pi$. If this is true for {\it both} of the sources in the geometry -- situated at $r = 0$ and $r = r_e$ -- then integrating the first of eqs.~\pref{newtotderivsr} between the two branes implies
\be \label{newtotderivsrx}
  3 \zeta H^2 \int_0^{r_e} \exd r \; a W^2 = W_e^4 (w_e + \tfrac12 \, q_e) \left( \frac{a_e}{r_e} \right) -  W_0^4 (w_0 + \tfrac12 \, q_0) \left( \frac{a_0}{r_0} \right)   =  0
   \,,
\ee
and so $H=0$, implying the maximally symmetric 4D geometry must be flat. Because \pref{wissmall} implies $w_b$ is always second-order in $\cL_b$ 4D curvature within this weak-coupling regime nonzero 4D curvature at linear order in $\cL_b$ requires a coupling between sources and $\varphi$. 

At second order in $\cL_b$ a nontrivial coupling to $\chi$ can suffice to obtain nonzero 4D curvature even if there is no direct coupling to $\varphi$, provided $\chi$ couples differently to the two sources. In this case solving the Kasner condition \pref{einsteinconstraintU2} for given $s_b$ allows nonzero $w_b \simeq s_b^2/16$ and so \pref{newtotderivsrx} can be nonzero at second order provided $s_0 \neq s_e$.

\subsubsection{Non-Kasner sources}

Conversely the non-Kasner asymptotic solutions whose powers are given by \pref{nonKasnerpowers} cannot describe weakly coupled sources because $s_b$, $q_b$ and $w_b$ are not perturbatively close to zero. In this case we must go back to the matching conditions \pref{scalarmatchingNS2phiw} through \pref{thetathetamatching2w}. In this case any mismatch beween the explicit $\epsilon$-dependence of the left-hand side and the $\epsilon$-dependence of the fields appearing in $\cL_b$ must be interpreted as indicating the need for an implicit $\epsilon$-dependence for the effective couplings\footnote{It is noteworthy that no $\epsilon$-dependent couplings were required when matching to D-branes in the stringy cosmic string solutions described in \cite{Bayntun:2009im}.} appearing in $\cL_b$.

This type of solution actually arises in the numerical solutions described above. For the plots shown all of the powers $q_b$, $s_b$, $w_b$ and $\alpha_b$ are chosen positive near $r = 0$ which implies $L_b$ for this source must be a growing function of both $\chi$ and $\varphi$. The solution obtained by integrating the field equations then predicts the asymptotic form near the other source and as shown in the plots, and these predict that only the sign of $\partial_r \varphi$ is opposite near the other source while all of the other derivatives do not change sign (provided one is careful to take the derivatives in the direction moving away from the source). This provides an indirect constraint on the form that $L_b$ must take and in particular requires the sources to couple to $\varphi$ with $L_b$ having opposite sign derivatives as $\varphi$ is varied. 

\subsubsection{D-brane charges}

A great benefit of knowing the F-theory pedigree of the 6D equations is that it allows a simple identification of whether the sources of the singularities can be well-known objects like D3 branes (or more generally bound states of D-branes). This can be done by testing whether the sources carry D3 or D7 charges, which in the current language corresponds to there being singularities in the fields to which these charges couple. Tracking through to the higher dimensions shows that D3 brane charge can be identified as a near-source singularity of the field $\Phi$ that is dual to the IIB 4-form $C_{(4)}$. D7 charges can similarly be read from the asymptotic behavior of the 10D 0-form $C_{(0)}$ that is a component of the complex string dilaton field.\footnote{Identifying D7 charges was done in \cite{Bayntun:2009im} for the flat solutions of 6D supergravity in which the precise D7 tension was obtained using PPEFT methods and agreeing with known results. Furthermore, in the 6D F-theory solutions found in \cite{Grimm:2013fua} were extended to non-trivial dilaton values by modifying slightly the metric ansatz. In both cases, the standard F-theory configuration with 24 singularities corresponding to D7 branes at weak coupling were reproduced. Notice that there are at least two types of D7 branes. Those wrapping the full $B_4$ base  and those wrapping a 2-cycle of $B_4$ and the two extra dimensions from going from 6D to 4D.} By this measure the solutions we find here have zero D3 and D7 charges because the fields $\Phi$ and $\tau$ are nonsingular at the source positions. We leave for a future publication the generalization of our solutions to include sources with nonvanishing D3/D7 charges by exploring solutions with non-trivial $\tau $ and $\Phi$ configurations. 


\section{Conclusions}
\label{Sec:Conclusions}
\begin{quote}
\rightline{{\it It's tough to make predictions, especially about the future.}}
\end{quote}

We outline in this paper a systematic  approach to obtain classical 4D maximally symmetric solutions of string theory: Minkowski, AdS or dS. Furthermore, we achieve these solutions without relying on quantum or $\alpha'$ corrections and therefore have better computational control as long as the corrections are subdominant and can be consistently neglected.

Our construction relies on several well-established tools developed over the past 20 years:
\begin{itemize}
\item The general classes of analytic and numerical Minkowski, AdS and dS solutions of gauged 6D supergravity.
\item The development of EFT techniques to describe the back-reaction of brane-like objects, tested with applications to brane and atomic systems.
\item The F-theory derivation of  gauged 6D supergravity including explicit ${\mathcal N}=1$ supersymmetric solutions reproducing known results from direct 4D compactifications.
\end{itemize}

Combining these three ingredients is relatively straightforward but far from trivial since the system of equations to be considered once the F-theory derivation of 6D supergravity is included, requires more scalar fields and their corresponding field equations. It is worth emphasizing that, contrary to Calabi-Yau compactifications, compactifications of 6D supergravity are much simpler, allowing for explicit metric solutions for the extra-dimensional geometry. Furthermore, the scalar fields do not take homogeneous values and instead have non-trivial profiles in the extra dimensions that can be explicitly computed, albeit so far only numerically.

The main point is that addressing simultaneously the two challenges to obtain dS in string theory -- namely the classical no-go theorem and the Dine-Seiberg problem in 6D rather than 4D -- leads to solutions of the field equations of 6D gauged supergravity derived from F-theory corresponding to 4D Minkowski, AdS and dS spacetimes. The solutions generically have co-dimension two singularities that have brane-like properties that can (but need not) correspond to positive-tension objects and the challenge of finding a full microscopic understanding of these extended objects remains an interesting open question. 

Similar singularities arise for the solutions found in \cite{Cordova:2018dbb} for massive IIA supergravity and applying the formalism of PPEFT to those solutions may also provide a better understanding how to interpret the properties of the sources to which they point. Furthermore, general studies of non-supersymmetric strings with runaway potentials have been made (see for instance \cite{Mourad:2024dur} and references therein) with similar properties in terms of brane-like singularities.  

We emphasize that from the EFT point of view this leaves us no worse off than when computing nuclear complications to atomic energy levels. One can compute the atomic effects of various nuclear moments without knowing the full  complicated structure of the atomic nucleus including all the details of strong interactions, confinement, etc. It is EFT techniques that allow us to compute the implications of finite-size nucleus size effects in precision calculations of the atomic energy levels by expanding in the small ratio of nuclear and atomic sizes. Similar techniques allow us to compute the back-reaction effects of the corresponding brane-like object at the singularities.

It is important to emphasize that having a monotonic potential where $V'=0$ need not be a bug and instead might be a feature. It forbids maximally symmetric solutions in the higher dimensions but does {\it not} exclude maximally symmetric solutions in lower dimensions, with non-trivial profiles for the scalar fields in the compactified dimensions. The solutions we find here are likely just the tip of the iceberg for a larger class of solutions of this type.

In the broader perspective, our work fits with the generic structure of the string landscape in which vacua with all signs of the cosmological constant are present. But unlike for Minkowski and AdS, explicit realization of dS backgrounds have been more challenging to obtain. We believe the results of this article provide progress in this direction and should help add to the accumulated evidence to the existence of dS solutions in string theory (and if they do not, why they do not). Furthermore, since our set up provides dS solutions within a chiral theory, it might be plausible to use it as a starting point for a search of realistic models in 4D. In particular our construction naturally fits with the anisotropic compactifications described in  \cite{Cicoli:2011yy} in which two extra dimensions can be much larger than the other extra dimensions providing an interesting hierarchy of scales in 4D.

Although the existence classical de Sitter solutions to the equations we solve is incontrovertible, several points remain open. These include the description of the full F-theory solutions with all moduli stabilized for concrete models, a better understanding of the singularities, a rigorous  proof of the stability of the solutions and so on. Furthermore, a natural next step after obtaining dS solutions is to look for time-dependent solutions giving rise to inflation starting not from a 4D effective action but from the full 10D equations which should be under reach following the work of \cite{vanNierop:2011di}. We believe this to be only  a first step towards a fully-fledged realistic classical dS solution from a concrete string construction. We leave the study of the scenarios 2 and 3 of section \ref{Sec:NewWave} for a follow-up publication \cite{followup} and hope to address some of the other questions  in the near future.

\section*{Acknowledgements}

We thank Yogi Berra for providing the quotes at the start of each section. We thank David Andriot, Thomas Grimm, Miguel Montero, Chris Pope, Alessandro Tomasiello  and Roberto Valandro for useful conversations and Susha Parameswaran for collaboration on an unsuccessful effort in this direction many years ago. CB's research was partially supported by funds from the Natural Sciences and Engineering Research Council (NSERC) of Canada. Research at the Perimeter Institute is supported in part by the Government of Canada through NSERC and by the Province of Ontario through MRI.  The work of FM is supported by a Stephen Hawking Fellowship.
The work of FQ was supported by STFC consolidated grants ST/P000681/1, ST/T000694/1.  FQ thanks Perimeter Institute for the kind hospitality, within the distinguished visiting research chair (DVRC) program, during the early stages of this project  and the CERN theory group for the excellent working conditions during the final stages of the project. This work is partly supported by COST (European Cooperation in Science and Technology) Action COSMIC WISPers CA21106.

\end{document}